\documentclass[twocolumn]{aastex63}
\usepackage{bm}
\usepackage{threeparttable}
\usepackage{morefloats}
\usepackage{CJK}
\usepackage{amsmath}
\usepackage{enumerate}
\usepackage[shortlabels]{enumitem}
\usepackage{mathrsfs}

\def\bfnabla{{\mbox{\boldmath $\nabla$}}}

\def\mbh{M_{\rm{BH}}}

\def\tnu{\tilde{\nu}}

\def\mprime{m^{\prime}}

\renewcommand\bv{{\mbox{\boldmath $v$}}}
\newcommand\bb{{\mbox{\boldmath $B$}}}
\newcommand\bP{{\mbox{\boldmath $P$}}}
\newcommand\bn{{\mbox{\boldmath $n$}}}

\newcommand\bF{{\mbox{\boldmath $F$}}}

\newcommand\Crat{{\mathbb{C}}}
\newcommand\Prat{{\mathbb{P}}}
\def\<{\,\langle\langle}
\def\>{\,\rangle\rangle}

\shorttitle{Multi-group RT Algorithm}
\shortauthors{Jiang}

\begin{document}
\begin{CJK*}{UTF8}{gbsn}

\title{Multi-group Radiation Magneto-hydrodynamics based on  Discrete Ordinates including Compton Scattering}

\correspondingauthor{Yan-Fei Jiang}
\email{yjiang@flatironinstitute.org}

\author[0000-0002-2624-3399]{Yan-Fei Jiang(姜燕飞)}
\affiliation{Center for Computational Astrophysics, \\
Flatiron Institute, \\
New York, NY 10010, USA}

\begin{abstract}
We present a formulation and numerical algorithm to extend the scheme for 
grey radiation magneto-hydrodynamics (MHD) developed by \cite{Jiang2021} to include 
the frequency dependence via the multi-group approach. The entire frequency space can be divided into arbitrary number of groups in the lab frame, and we follow the time dependent evolution of frequency integrated specific intensities along 
discrete rays inside each group. Spatial transport of photons is done in the lab frame while all the coupling terms are solved in the fluid rest frame. Lorentz transformation is used to connect different frames.  Radiation transport equation is solved fully implicitly in time while the MHD equations are evolved explicitly so that time step is not limited by the speed of light. A finite volume approach is used for transport in both spatial and frequency spaces to conserve radiation energy density and momentum. The algorithm includes photon absorption, electron scattering as well as Compton scattering, which is calculated by solving the Kompaneets equation. The algorithm is accurate for a wide range of optical depth conditions and can handle both radiation pressure and gas pressure dominated flows. It works for both Cartesian and curvilinear coordinate systems with adaptive mesh refinement. We provide a variety of test problems including radiating sphere, shadow test, absorption of a moving gas, Bondi type flows as well as a collection of test problems for thermal and bulk Compton scattering. We also discuss examples where frequency dependence can make a big difference compared with the grey approach.

\end{abstract}


\section{Introduction}
\label{sec:introduction}
Radiation magneto-hydrodynamic (MHD) simulations play an important role for theoretical studies of many astrophysical problems. They can be used to determine thermal properties of the system with the dynamical evolution together, which can be directly connected to many observables such as lightcurves and spectra. In many cases, momentum  and energy exchanges between photons and plasma provide the dominant pressure as well as heating and cooling source and radiation transport (RT) is essential in the simulations \citep[see the review][and references therein]{Teyssier2015}. Properties of photons are described by the fundamental quantity, specific intensity, which can vary with time, space, photon propagation direction, frequency as well as polarization in general. However,  different levels of approximations are typically adopted to simplify the calculation.  One simplification is to take the moments of specific intensities and make certain assumptions to close the moment equations such as Flux-limited diffusion (FLD) \citep[e.g.,][]{LevermorePomraning1981,TurnerStone2001,Krumholzetal2007,Holstetal2011,Zhangetal2011,Zhangetal2013,Kuiperetal2020}, the M1 approach \citep{DubrocaFeugeas1999,Ponsetal2000,Gonzalezetal2007,SkinnerOstriker2013,Sadowskietal2013,McKinneyetal2013,Skinneretal2019,AnninosFragile2020}, or using variable Eddington tensors as obtained by solving another set of \emph{time-independent} transport equation based on short characteristics \citep{Stoneetal1992,HayesNorman2003,Jiangetal2012,Davisetal2012,Asahinaetal2020,Menonetal2022}. Another important simplification is that the evolved radiation field is typically integrated over the frequency space and different types of averaged opacities are adopted in the calculation. 

In cases when radiation spectrum cannot be described by a simple formula such as blackbody, or the traditional averaged opacities (such as Rosseland mean and Planck mean) cannot be used, frequency dependence also needs to be included in RT calculations. For example, UV and infrared photons are typically emitted by different sources and they can have very different interactions with gas and dust particles. Near the photosphere where diffusion assumption cannot apply, even under the approximation of local thermal equilibrium (LTE), Rosseland mean opacities may not describe the momentum exchange between radiation and gas correctly. This is also the region where radiation spectrum can differ from blackbody significantly and frequency dependent RT is needed. 
There are several ways to include the frequency dependence of photons in RT calculations. The first straight forward approach is to solve the RT equation at discrete frequencies and the full spectrum is sampled with a certain number of frequency points \citep{Dydaetal2019}. This approach is typically taken by Monte Carlo RT calculations (see the recent review by \citealt{NoebauerSim2019} and references therein) and many ray-tracing type RT solvers that are based on short or long characteristics \citep[see, e.g.,][]{Mihalasetal1978,Hayeketal2010,Davisetal2012,Frostholmetal2018,Wunschetal2021}. 
Because radiation MHD calculations are typically much more expensive than MHD simulations and the cost will increase at least linearly with the number of frequency points, another commonly used approach to include the frequency dependence is multi-group \citep{MihalasMihalas1984,Vaytetetal2011,Zhangetal2013}, which divides the frequency space into a small number of groups. We then evolve the specific intensities which are frequency integrated inside each group. Properly weighted opacities (such as the Rosseland mean and Planck mean) are normally needed for each frequency bin. This approach can cover the frequency space continuously to allow a better energy conservation when photons are redistributed in the frequency space. This is also the approach we will take. It can be used to handle physical processes that have a smooth dependence of frequency such as Compton scattering. However, it will not be able to handle line transport easily without a large number of frequency groups. One special type of multi-group RT uses the so called opacity distribution function method, which divides the frequency space based on the ordering of opacity as a function of frequency \citep{Skartlien2000,Anushaetal2021,Witzkeetal2021}. It can be used to handle strong opacity variations  as a function of frequency efficiently. However, future development is still needed to account for the frequency change due to Doppler effects correctly.

Coherent Thomson scattering process is widely included in RT calculations, where photons and gas can only exchange momentum but not energy during each scattering process. However, in many astrophysical systems, incoherent Compton scattering, where photons and gas can exchange both energy and momentum, plays an important role to determine thermal properties of the plasma as well as spectrum properties of the radiation field. This includes thermal Compton with high temperature electrons (typically $10^6$ K or above), or bulk Compton due to large turbulent or bulk velocities of electrons. For example, the corona regions in accretion disks around black holes, where Compton upscattering of disk photons by high temperature or non-thermal electrons are believed to be the dominant mechanism to produce the commonly observed hard X-rays \citep{HaardtMaraschi1991,Svenssonetal1994}. In Gamma-ray bursts \citep{Thompson1994,Zrakeetal2019}, supernova shock breakout \citep{Chevalieretal2008,SuzukiShigeyama2010}, X-ray pulsars \citep{CaballeroWilms2012} and polar cap accretion onto neutron stars \citep{BaskoSunyaev1975,Zhangetal2022}, Compton scatterings are all believed to play an important role to determine the spectrum properties. The well known Sunyave-Zel'dovich effect is also  
just due to inverse Compton scattering of cosmic microwave background by the high energy electrons in galaxy cluster \citep{SunyaevZeldovich1970,Rephaeli1995}. 

Various numerical techniques have been developed to calculate the Comptonized spectrum. One common approach is Monte Carlo based methods\citep{Pozdnyakovetal1983,GoreckiWilczewski1984,Dolenceetal2009,SchnittmanKrolik2013,Ryanetal2015,Rothetal2022}, which are flexible and can handle the relativistic scattering kernel accurately. However, it is typically not efficient in the optically thick regime and it is also hard to include stimulated emission term, which is necessary to get the blackbody spectrum in thermal equilibrium. Deterministic methods by solving specific intensities along discrete rays based on either short or long characteristics have also been developed \citep{Hubenyetal2001,Psaltis2001,Narayanetal2016}, which typically solve the Kompaneets equation for Compton scattering. Most of these codes solve the RT equation 
in the post processing way, which means properties of the plasma are froze while the radiation spectrum is calculated. In this paper, we want to develop a scheme that is able to couple the Compton process with the full time dependent evolution of radiation magneto-hydrodynamic equations together. The scheme needs to work accurately and efficiently for any optical depth regime. Stimulated emission needs to be included so that we can recover the blackbody spectrum in thermal equilibrium. A numerical scheme that is able to accurately follow the time dependent evolution of radiation and gas is also fundamentally different from post processing calculations, as we need to ensure total energy and momentum conservation accurately. We will show in the following sections that the algorithm developed here satisfies all these constrains. 

In \cite{Jiang2021}, we developed an implicit scheme that solves the time dependent, frequency integrated RT equation directly based on discrete ordinates. Here we extend the algorithm to include the frequency dependence via the multi-group approach. The extended algorithm has maintained all the properties of the original scheme but also allows frequency dependent opacities with very flexible decomposition of the frequency space in a conservative way. This paper is organized as follows.  In section \ref{sec:eq}, we describe how the frequency groups are transformed in different frames via Lorentz transformation, as well as the full multi-group radiation MHD equations to solve. Numerical implementations of the algorithm are described in section \ref{sec:num}. In section \ref{sec:test}, series of tests are described to demonstrate the multi-group capability of the new algorithm. Performance of the algorithm and future extensions are discussed in section \ref{sec:discuss}.

\section{Equations}
\label{sec:eq}
We will extend the frequency integrate RT equation based on discrete ordinates as described in \cite{Jiang2021} by including multiple frequency groups. This require changes to 
source terms and transformation of specific intensities as well as frequency space in different frames as shown in the following sections. One important property we always keep is that if only one frequency group is used, 
the formula will be automatically reduced to grey RT. 

\subsection{Frame Transformation For Multi-group Specific Intensities}

The fundamental quantity we use to describes the radiation field is the lab frame specific intensity $I_{\nu}$, which is a function of time $t$, spatial locations $(x,y,z)$, angular direction $\bn$ and monochromatic 
frequency $\nu$. We divide the frequency space into fixed $N_f$ bins $[0,\nu_1),[\nu_1,\nu_2),......,[\nu_{N_f-2},\nu_{N_f-1}),[\nu_{N_f-1},\infty)$ for $N_f\geq 2$. If $N_f=1$, it will be just the whole frequency space $[0,\infty)$. 
In each frequency 
bin $[\nu_f,\nu_{f+1})$, we define the frequency integrated specific intensity as 
\begin{eqnarray}
I_f\equiv \int_{\nu_f}^{\nu_{f+1}} I_{\nu} d\nu.
\end{eqnarray}
If we only have one frequency bin, this is reduced to the frequency integrated grey specific intensity. For the multi-group approach, we will evolve $I_f$ as the discretized representation of $I_{\nu}$ . 

\begin{figure}[htp]
	\centering
	\includegraphics[width=1.0\hsize]{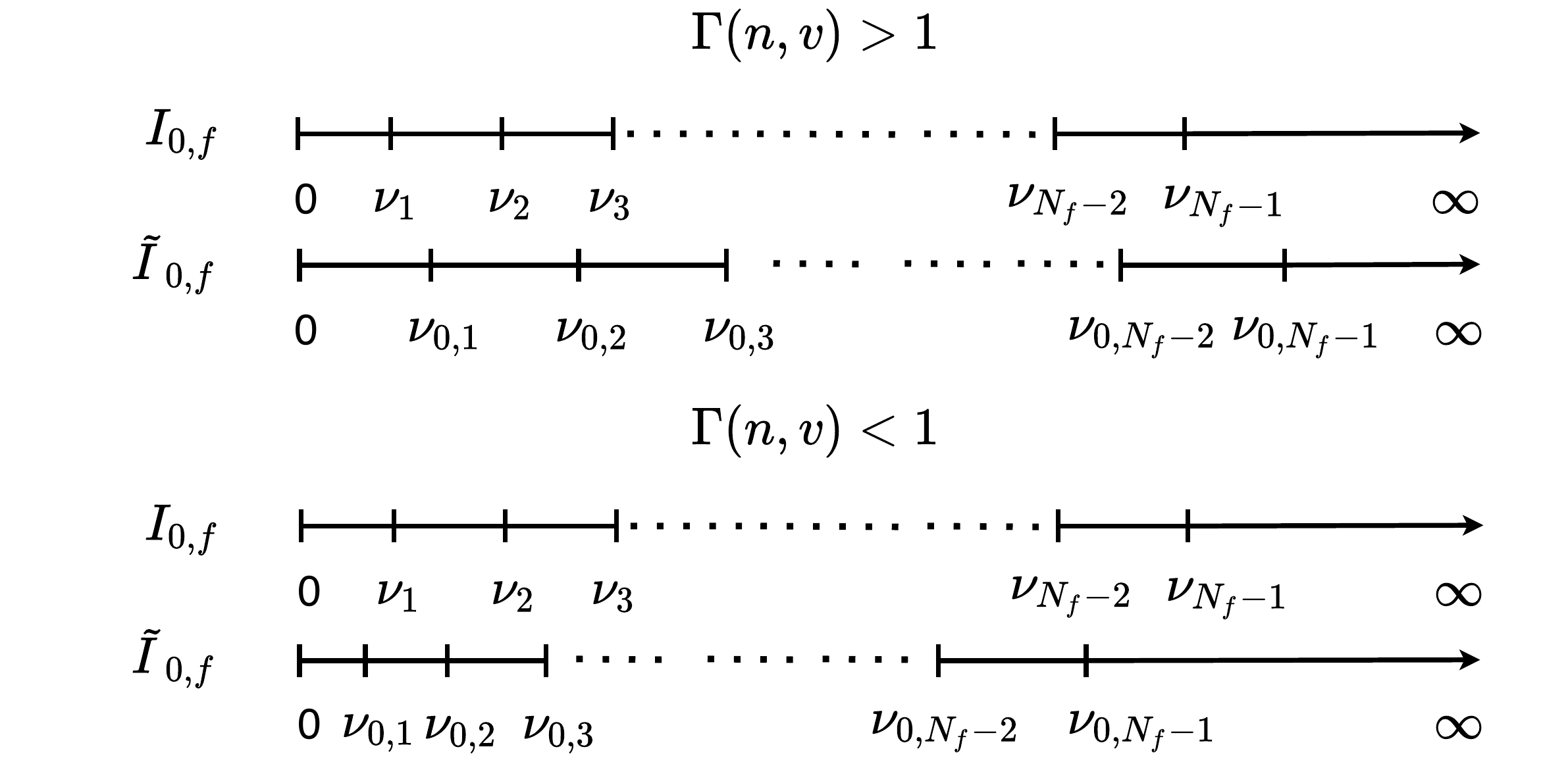}
	\caption{Schematic illustrations of mapping co-moving frame specific intensities between two 
	different sets of frequency grids.  The top and bottom panels show two cases when the Lorentz transformation 
   factor $\Gamma(n,v)$ is larger and smaller than 1 respectively. }
	\label{fig:fre_map}
\end{figure}

The numerical scheme will require transformation of specific intensities between lab frame and co-moving frame in order to solve the coupling terms between radiation and gas. Frequency $\nu$ in the lab frame is related to the 
frequency in the co-moving frame $\nu_0$ as 
\begin{eqnarray}
	\nu_{0}= \gamma\left(1-\bn\cdot\bv/c\right) \nu\equiv \Gamma(\bn,\bv) \nu,
	\label{eq:fre_transform}
\end{eqnarray}
where $\bv$ is the flow velocity, $c$ is the speed of light and $\gamma$ is the corresponding Lorentz factor. The angle 
in the co-moving frame $\bn^{\prime}$ is related to the angle in the lab frame $\bn$ as \citep{MihalasMihalas1984}
\begin{eqnarray}
	\bn^{\prime}=\frac{1}{\gamma\left(1-\bn\cdot\bv/c\right)}\left[\bn-\gamma\frac{\bv}{c}\left(1-\frac{\gamma}{\gamma+1}\frac{\bn\cdot\bv}{c}\right)\right],
	\label{eqn:cm_ang}
\end{eqnarray}
which is independent of frequency. This means once we choose the same set of angles for all the frequency bins in the lab frame, the angular grid will still be the same for different frequency bins in the co-moving frame. However, frequency grid $\nu_f$, which is independent of angles, will end up being different co-moving frame frequency grids $\nu_{0,f}$ for different angle $\bn$. 

Specific intensity $\tilde{I}_{0,f}(\bn^{\prime})$ in the co-moving frame is related to $I_f(\bn)$ as
\begin{eqnarray}
\tilde{I}_{0,f}(\bn^{\prime})&\equiv &\int_{\nu_{0,f}}^{\nu_{0,f+1}} I_{\nu_0} d\nu_0=\int_{\nu_{0,f}}^{\nu_{0,f+1}} I_{\nu}\left(\frac{\nu_0}{\nu}\right)^3d\nu_0\nonumber\\
&=&\Gamma^4\left(\bn,\bv\right)\int_{\nu_{f}}^{\nu_{f+1}} I_{\nu} d\nu\equiv \Gamma^4(\bn,\bv) I_f.
\label{eq:I_transform}
\end{eqnarray}
Here we have used the fact that $I_{\nu}/\nu^3$ is Lorentz invariant and $\nu_{0,f}=\Gamma(\bn,\bv)\nu_f, \nu_{0,f+1}=\Gamma(\bn,\bv)\nu_{f+1}$. 

It is very inconvenient to use $\tilde{I}_{0,f}(\bn^{\prime})$ in the co-moving frame, as the frequency integral range $[\nu_{0,f},\nu_{0,f+1})$ will vary as a function of angle $\bn$.  We define a new set of specific intensities in the co-moving frame as
\begin{eqnarray}
I_{0,f}(\bn^{\prime})=\int_{\nu_{f}}^{\nu_{f+1}} I_{\nu_0} d\nu_0.
\label{eq:cm_intensity}
\end{eqnarray}
Both $I_{0,f}$ and $\tilde{I}_{0,f}$ are discretized representations of the same co-moving frame specific intensities $I_{\nu_0}$ in two different sets of frequency grids that cover the whole frequency range $[0,\infty)$. We define two sets of frequency mapping operators $\mathcal{M}_{\nu}$ and $\mathcal{M}^{-1}_{\nu}$ such that
\begin{eqnarray}
I_{0,f}=\mathcal{M}_{\nu}\left(\tilde{I}_{0,f}\right), \tilde{I}_{0,f}=\mathcal{M}^{-1}_{\nu}\left(I_{0,f}\right).
\label{eq:fre_mapping}
\end{eqnarray}
Our implementation of the two operators will guarantee that $\mathcal{M}^{-1}_{\nu}\mathcal{M}_{\nu}$ is the identity operator to the machine precision. 

 The forward mapping operator $\mathcal{M}_{\nu}$ is illustrated in Figure \ref{fig:fre_map}. For each frequency bin $[\nu_f,\nu_{f+1})$, we first find the location in the frequency grid $\nu_{0,f}$ such that $\nu_{0,f_L}\leq\nu_f<\nu_{0,f_L+1}$ and $\nu_{0,f_R}\leq\nu_{f+1}<\nu_{0,f_R+1}$,  where $f_L$ and $f_R$ represent the bin numbers that $\nu_f$ and $\nu_{f+1}$ are located.  
Then $I_{0,f}$ is simply the sum of $\tilde{I}_{0,f}$ for all frequency bins between $\nu_{0,f_L+1}$ and $\nu_{0,f_R}$ (if $\nu_{0,f_R}>\nu_{0,f_L+1}$) plus a fraction of bins where $\nu_f$ and $\nu_{f+1}$ are located. For example, when $\Gamma(\bn,\bv)>1$, we determine the integral $\int_{\nu_f}^{\nu_{0,f_L+1}}I_{\nu_0}d\nu_0$ by assuming 
$I_{\nu_0}=\tilde{I}_{0,f_L}/(\nu_{0,f_L+1}-\nu_{0,f_L})$ at $\nu_0=(\nu_{0,f_L}+\nu_{0,f_L+1})/2$ and piecewise linear 
reconstruction of $I_{\nu_0}\in [\nu_{0,f_L},\nu_{0,f_L+1})$ in the same way as done for hydro variables in {\sf Athena++} \citep{Stoneetal2020}. 
However, if 
this will cause new extrema in the spectrum, we will instead assume $I_{\nu_0}$ is a constant in this bin. 
The integral $\int_{\nu_{0,f_R}}^{\nu_{f+1}} I_{\nu_0}d\nu_0$ is calculated in the same 
way based on $\tilde{I}_{0,f_R}$. Similar mapping is done in the case $\Gamma(\bn,\bv)<1$ as shown in Figure \ref{fig:fre_map}. The only special case is the last bin 
$[\nu_{N_f-1},\infty)$ when $\Gamma(\bn,\bv)<1$. In this case, 
we cannot split the last bin with the same method. Instead, we assume $I_{\nu_0}$ 
follows a blackbody spectrum at the temperature $T_{r,N_f-1}$, which is determined by requiring integral of the blackbody spectrum 
over the frequency range $[\nu_{0,N_f-1},\infty)$ is $\tilde{I}_{0,N_f-1}$. Different assumptions on the spectrum shape in this bin can also be used in principle. 
 Then we split $\tilde{I}_{0,N_f-1}$ into two 
integrals between $[\nu_{0,N_f-1},\nu_{N_f-1})$ and $[\nu_{N_f-1},\infty)$ using the same blackbody spectrum. 
In this way, the sum of $\tilde{I}_{0,f}$ and $I_{0,f}$ are identical after the mapping. 

The inverse mapping operator $\mathcal{M}_{\nu}^{-1}$ is only used after $\mathcal{M}_{\nu}$ is done. 
 If we use the same mapping scheme as illustrated in Figure \ref{fig:fre_map}, $\mathcal{M_{\nu}}^{-1}\mathcal{M}_{\nu}$ is generally not the identify operator.  In order to avoid this numerical diffusion, we first store the contributions from all groups of $\tilde{I}_{0,f}$ that contribute to each bin $I_{0,f}$ during the mapping process from $\tilde{I}_{0,f}$ to $I_{0,f}$, which means $I_{0,f}=\sum_{f_i} \tilde{I}_{0,f_i}$. We then calculate the ratio $R_{f_i}\equiv \tilde{I}_{0,f_i}/I_{0,f}$. When we need to map $I_{0,f}$ back to the frequency grid $\tilde{I}_{0,f}$, we assume the ratios $R_{f_i}$ are unchanged and 
$R_{f_i}I_{0,f}$ is added back to the frequency bin $\tilde{I}_{0,f_i}$. In this way, we can make sure 
$\tilde{I}_{0,f}=\mathcal{M}_{\nu}^{-1}\mathcal{M}_{\nu}(\tilde{I}_{0,f})$.  We find this is necessary to maintain steady state solutions for arbitrary spectrum shape.

\subsection{The Angular Grid}
\label{sec:angular_grid}
The two types of angular grids mentioned in \cite{Jiang2021} are all currently supported for this multi-group RT algorithm and the implementation is identical. We summarize the notations here for completeness, which will also be used in the following sections.  The first set of angles $\bn$ is defined with 
respect to fixed directions and they do not change with spatial location in the grid,  even if the coordinate systems are curvilinear \citep{Carlson1963,Davisetal2012}. The second set of angles is defined at fixed directions with respect to local coordinate axes in each cell. The direction vector $\bn$ itself will then change from cell to cell and additional terms are needed to account for this angular flux as described in section 3.2.4 of \cite{Jiang2021} following the formula given by \cite{DavisGammie2020}. This angular system can be very useful for simulations in spherical polar coordinate. In either angular grid, each angle $\bn$ has three components $\bn=(n_x,n_y,n_z)$ and the corresponding quadrature weights $w_n$ in the lab frame, which is determined by the angular volume $\bn$ represents. The commonly used  radiation energy density $E_r$, flux $\bF_r$ and pressure $\bP_r$ are not needed in our RT equation. But they are useful diagnostics and can be constructed by angular quadrature of the specific intensities in each frequency bin as
\begin{eqnarray}
	E_{r,f}&=&4\pi \sum_{n=0}^{N-1} I_f(\bn) w_n, \nonumber\\ 
	\bF_{r,f}&=&4\pi c \sum_{n=0}^{N-1}I_f(\bn) \bn w_n, \nonumber\\
	\bP_{r,f}&=&4\pi \sum_{n=0}^{N-1}I_f(\bn) \bn\bn w_n,
\end{eqnarray}
where $N$ is the total number of angles per cell. Similar calculations can be done to get the moments in the co-moving frame except that we need to use the angles in the co-moving frame $\bn^{\prime}$ and the co-moving frame quadrature weight $w_n^{\prime}=\Gamma^{-2}w_n$ for co-moving frame specific intensities $I_{0,f}(\bn^{\prime})$. Frequency integrated moments of the radiation field are simply 
\begin{eqnarray}
	E_{r}=\sum_{f=0}^{N_f-1} E_{r,f}; 
	\bF_{r}=\sum_{f=0}^{N_f-1} \bF_{r,f};	
	\bP_{r}=\sum_{f=0}^{N_f-1} \bP_{r,f}. 
\end{eqnarray}

 \subsection{Equations for Multi-group RT}
 The lab frame monochromatic RT equation can  be written as \citep{MihalasMihalas1984}
 \begin{eqnarray}
 	\frac{\partial I_{\nu}}{\partial t}+c\bn\cdot\bfnabla I_{\nu}=c\left(\eta_{\nu}-\chi_{\nu}I_{\nu}\right),
 \end{eqnarray}
where $\eta_{\nu}$ and $\chi_{\nu}$ are the monochromatic emissivity and opacity. We integrate the left and right hand 
sides over frequency for each bin $[\nu_f,\nu_{f+1})$ to get the equation we actually solve for frequency integrated specific 
intensities in each bin
 \begin{eqnarray}
 	\frac{\partial I_{f}}{\partial t}+c\bn\cdot\bfnabla I_{f}=c\left(\eta_{f}-\chi_f I_{f}\right),
 	\label{eq:full_rt}
 \end{eqnarray}
 where $\eta_f\equiv \int_{\nu_f}^{\nu_{f+1}} \eta_{\nu} d\nu$ and $\chi_f\equiv \int_{\nu_f}^{\nu_{f+1}} \chi_{\nu} I_{\nu} d\nu/I_f$.
Expressions for the source terms in the co-moving frame can be determined in a similar way as done for grey RT  \citep{Jiang2021}. 
We apply the frame transformation operator $\Gamma^4(\bn,\bv)$ to both sides of equation \ref{eq:full_rt} 
and then the frequency mapping operator $\mathcal{M}_{\nu}$. We assume scattering, absorption and emission are isotropic 
in the fluid rest frame. All these terms are integrated over frequency for each group in the co-moving frame. 
The resulting source terms can be written as 
\begin{eqnarray}
	\frac{\partial I_{0,f}}{\partial t}&=&c\Gamma\left[\right. \rho\kappa_s(J_{0,f}-I_{0,f})
	+\rho\kappa_{a,f}\left(\varepsilon_{0,f}-I_{0,f}\right) \nonumber\\
	&+&\rho(\kappa_{p,f}-\kappa_{a,f})\left(\varepsilon_{0,f}-J_{0,f}\right)\left.\right] \nonumber\\
	&=&c\Gamma\left[\right.\rho\left(\kappa_s+\kappa_{a,f}\right)\left(J_{0,f}-I_{0,f}\right)\nonumber\\
	&+&\rho\kappa_{p,f}\left(\varepsilon_{0,f}-J_{0,f}\right)\left.\right].
\label{eq:source}
\end{eqnarray}

Here $\varepsilon_{0,f}$ is the frequency integrated thermal 
emissivity for each frequency bin. In principle, $\varepsilon_{0,f}$ can be any function of local gas and radiation properties. Our default assumption is 
local thermal equilibrium so that $\varepsilon_{0,f}$ is determined by the gas temperature $T$ as 
\begin{eqnarray}
\varepsilon_{0,f}&=&\frac{1}{4\pi}\int_{\nu_f}^{\nu_{f+1}} B(\nu, T) d\nu,\nonumber\\
&=&\frac{1}{4\pi}\int_{h\nu_f/k_BT}^{h\nu_{f+1}/k_BT} B(\nu_T) d\nu_T,\nonumber\\
B(\nu,T)&=&\frac{8\pi h\nu^3}{c^3}\frac{1}{\exp\left(h\nu/k_BT\right)-1},\nonumber\\
B(\nu_T)&=&\frac{15a_rT^4}{\pi^4}\frac{\nu_T^3}{\exp(\nu_T)-1}.
\label{eq:emissivity0}
\end{eqnarray}
We have defined the dimensionless frequency as $\nu_T\equiv h\nu/k_BT$ and $a_r$ is the radiation constant while $k_B$ is the Boltzmann constant. It can be easily checked that if we only have one frequency bin, $\varepsilon_{0,f}$ is reduced to the normal thermal emission term $\varepsilon_{0,f}=a_rT^4/\left(4\pi\right)$. The mean radiation energy density in each frequency bin $J_0$ is just the angular quadrature of the specific intensities as
\begin{eqnarray}
	J_{0,f}=\int I_{0,f}d\Omega_0,
\end{eqnarray}
where $\Omega_0$ is the weight of angular quadrature  in the co-moving frame. This can be done because co-moving frame specific intensities $I_{0,f}$ share the same frequency grid $\nu_f$ for different angles. 
Scattering opacity $\kappa_s$ is typically taken to be the electron scattering value while the Rosseland mean ($\kappa_{a,f}$) and Planck mean ($\kappa_{p,f}$) absorption opacities in each frequency bin are defined as
\begin{eqnarray}
\frac{1}{\kappa_{a,f}}&\equiv&\frac{1}{\int_{\nu_{f}}^{\nu_{f+1}}B_T(\nu,T) d\nu}\int_{\nu_f}^{\nu_{f+1}}\frac{1}{\kappa_{a,\nu}}B_T(\nu,T) d\nu, \nonumber\\
\kappa_{p,f}&\equiv& \frac{1}{\int_{\nu_{f}}^{\nu_{f+1}}B(\nu,T) d\nu}\int_{\nu_f}^{\nu_{f+1}}\kappa_{a,\nu}B(\nu,T) d\nu, \nonumber\\
B_T(\nu,T)&\equiv& \frac{\partial B(\nu,T)}{\partial T}.
\label{eq:ross_planck_opacity}
\end{eqnarray}
Here $\kappa_{a,\nu}$ is the absorption opacity at the monochromatic frequency $\nu$. If the integral is taken between $[0,\infty)$, they are the commonly used Rosseland mean and Planck mean opacities in grey RT. The opacities are defined in the co-moving frame but the integral is done for the frequency grid $\nu_f$.

These source terms only include isotropic scattering and radiation energy density does not change during the scattering process. Additional terms need to be added to account for Compton scattering, which is described in the following section \ref{sec:eq_compt}. We have also approximated the specific intensity weighted opacities with the Rosseland mean and Planck mean values \citep{Pomraning1973}, as $\chi_f$ is generally unavailable. It can be easily confirmed that if we integrate the specific intensities over all angles in the co-moving frame, $\kappa_s+\kappa_{a,f}$ will determine the momentum coupling between photons and gas while only $\kappa_{p,f}$ will be responsible for the thermal coupling as expected. Notice that if we can use large enough number of frequency bins to resolve the frequency dependence of opacities, $\kappa_{a,f}$ and $\kappa_{p,f}$ will both approach $\kappa_{a,\nu}$ and the term $(\kappa_{p,f}-\kappa_{a,f})\left(\varepsilon_{0,f}-J_{0,f}\right)$ will go away.

\subsection{Kompaneets Equation for Compton Scattering}
\label{sec:eq_compt}

In the limit that $h \nu,\ k_BT \ll m_e c^2$, Kompaneets equation \citep{Kompaneets1957} can be used to describe the energy exchange between photons and gas via Compton scattering. Although full angular dependent Compton scattering kernel can be used as in many Monte Carlo calculations \citep{Pozdnyakovetal1983,Psaltisetal1997,Hubenyetal2001}, it significantly complicates the equation and makes the scheme much more expensive. Instead, we will assume that the energy exchange due to Compton scattering is uniformly distributed over all angles in the co-moving frame of the gas, which should be a good assumption when radiation flux is much smaller than $cE_r$. Therefore the momentum exchange between photon and gas is described by the usual scattering term, which is already included in equation \ref{eq:source}. 

The Kompaneets equation evolves the photon occupation number $n_{\nu}$, which is related to the monochromatic radiation energy density $J_{\nu}$ as 
\begin{eqnarray}
	n_{\nu}=\left(\frac{c^3}{8\pi h}\right)\frac{J_{\nu}}{\nu^3},
	\label{eq:jton}
\end{eqnarray}
Evolution of $n_{\nu}$ due to Compton scattering follows this equation
\begin{equation}
	\frac{1}{c\rho\kappa_{\rm es}}\frac{\partial n_{\nu}}{\partial t}=\frac{k_BT}{m_ec^2}\frac{1}{\nu_T^2}\frac{\partial}{\partial \nu_T}\left(\nu_T^4\left[\frac{\partial n_{\nu}}{\partial \nu_T}+n_{\nu}(n_{\nu}+1)\right]\right),
\end{equation}
where $\kappa_{\rm es}$ is the frequency independent electron scattering value. 
 We assume gas temperature is fixed while this equation is solved. Then we can change the independent variable from $\nu_T$ to the dimensionless frequency $\tnu\equiv h\nu/k_BT_0$, which is independent of gas temperature $T$, as 
\begin{equation}
	\frac{1}{c\rho\kappa_{\rm es}}\frac{\partial n_{\tnu}}{\partial t}=\frac{1}{T_e}\frac{1}{\tnu^2}\frac{\partial}{\partial \tnu}\left(\tnu^4\left[T\frac{\partial n_{\tnu}}{\partial \tnu}+n_{\tnu}(n_{\tnu}+1)\right]\right),
	\label{eq:compton}
\end{equation}
where we have defined $T_e\equiv m_e c^2/k_BT_0$ and gas temperature $T$ is scaled with our fiducial temperature unit $T_0$. 

Numerically, we will implement a conservative format of the above equation for total photon numbers $N=\int  n_{\tnu} \tnu^2 d\tnu$ as
\begin{eqnarray}
	\frac{T_e}{c\rho\kappa_{\rm es}}\frac{\partial \left(\tnu^2n_{\tnu}\right)}{\partial t}=\frac{\partial}{\partial \tnu}\left[ \tnu^4 F(n_{\tnu})\right],
\end{eqnarray}
which describes the process that photons are just transported between different frequency bins with the flux $\tnu^4 F(n_{\tnu})$ and 
\begin{eqnarray}
	F(n_{\tnu})\equiv T\frac{\partial n_{\tnu}}{\partial \tnu}+n_{\tnu}\left(n_{\tnu}+1\right).
\label{eq:compt_fn}
\end{eqnarray}
The first term represents diffusion in the frequency space while the non-linear term $n_{\tnu}^2$ is due to stimulated emission. 
In steady state, $F(n_{\tnu})$ needs to be zero for all frequency bins and the general Bose-Einstein distribution $n_{\tnu}=1/\left[\lambda \exp(\tnu/T)-1\right]$ satisfies this condition. Blackbody distribution is just a special case with $\lambda=1$.
Notice that if the stimulated emission term $n_{\tnu}^2$ is not included, the steady state spectrum will take the Wien distribution $n_{\tnu}\propto \exp{\left(-\tnu/T\right)}$.

The change of mean radiation energy density due to Compton process can be determined by integrating the left and right hand sides of equation \ref{eq:compton} with $\tnu^4 d\tnu$. It can be shown that if the gas temperature $T$ equals the Compton temperature $T_c$ (in unit of $T_0$) of the radiation field as defined in the following way
\begin{eqnarray}
	T_c\equiv \left(\int_0^{\infty}n_{\tnu}\tnu^3 d\tnu\right)^{-1}\left[\frac{1}{4}\int_0^{\infty} n_{\tnu} \tnu^4 d\tnu \right. \nonumber\\
	+\left. \frac{\pi^4}{60}\int_0^{\infty}n_{\tnu}^2\tnu^4 d\tnu	\right],
\label{eq:compton_tem}
\end{eqnarray}
there will be no change of the mean radiation energy density for any spectrum.

In section \ref{sec:num_compton}, we will describe how this Kompaneets equation is used to determine the change of $I_{0,f}$ numerically. In principle, equation \ref{eq:compton} can be converted to an equation that evolves $J_{\nu}$ directly as done in  \cite{Hubenyetal2001}. However, we prefer the current format to evolve photon number density as it has a nice conservation property for the Compton process.

\subsection{Full Equations for Multi-group Radiation MHD}
Radiation energy and momentum source terms for the gas are simply the sum of contributions from photons in all frequency bins. In the lab frame, the total energy and momentum of gas and radiation are conserved. For completeness, we summarize the full multi-group radiation MHD equations as
\begin{eqnarray}
	\frac{\partial\rho}{\partial t}+\bfnabla\cdot(\rho \bv)&=&0, \nonumber \\
	\frac{\partial( \rho\bv)}{\partial t}+\bfnabla\cdot({\rho \bv\bv-\bb\bb+{{\sf P}^{\ast}}}) &=&-\sum_{f=0}^{N_f-1} \bm{S}_f(\bP),\  \nonumber \\
	\frac{\partial{E}}{\partial t}+\bfnabla\cdot\left[(E+P^{\ast})\bv-\bb(\bb\cdot\bv)\right]&=&-\sum_{f=0}^{N_f-1} S_{f}(E),  \nonumber \\
	\frac{\partial\bb}{\partial t}-\bfnabla\times(\bv\times\bb)&=&0, 
		\label{eq:mhd}
\end{eqnarray}
Here $\rho, \bv, P$ and $\bb$ are density, flow velocity, gas pressure and the magnetic field respectively with the total pressure ${\sf P}^{\ast}\equiv(P+B^2/2){\sf I}$ ( ${\sf I}$ is
the unit tensor). Ideal MHD has been assumed for the induction equation. Total energy density is defined as 
\begin{eqnarray}
	E=E_g+\frac{1}{2}\rho v^2+\frac{B^2}{2},
\end{eqnarray}
where gas internal energy is $E_g=P/(\gamma_g-1)$ for the adiabatic index 
$\gamma_g \neq 1$. The gas temperature is calculated with 
$T=\left(\mu m_p/k_B\right)\left(P/\rho\right)$, where $\mu$ is the mean molecular weight and $m_p$ is the proton 
mass. Momentum and energy source terms due to photons in each frequency group are represented by $\bm{S}_f(\bP)$ and $S_{f}(E)$. 

The corresponding equations for specific intensities are
\begin{eqnarray}
	\frac{\partial I_f}{\partial t}&+&c\bn\cdot\bfnabla I_f=cS_f(I),\nonumber\\
	S_f(I)&\equiv& \Gamma^{-3}\mathcal{M}_{\nu}^{-1}\left[\right. \rho(\kappa_s+\kappa_{a,f})\left(J_{0,f}-I_{0,f}\right)\nonumber\\
	&+&\rho\kappa_{p,f}\left(\varepsilon_{0,f}-J_{0,f}\right)+S_c(I_f)\left.\right],\nonumber\\
	I_{0,f}&=&\mathcal{M}_{\nu}\Gamma^4 I_f,\nonumber\\
	S_f(E)&\equiv& 4\pi c\int S_f(I) d\Omega,\nonumber\\
	\bm{S}_f(\bP)&\equiv& 4\pi \int \bn S_f(I) d\Omega,
\label{eq:rad_mhd}
\end{eqnarray}
where $S_c(I_f)$ is the source term due to Compton scattering, which will be described in Section \ref{sec:num_compton}. The weight 
of angular quadrature in the lab frame $d\Omega$ is also normalized such that $\int d\Omega=1$.

This set of RT equations is basically the Newtonian version of the covariant RT equation derived by \cite{DavisGammie2020} 
after we integrate  specific intensities over frequency inside each frequency bin.  
Notice that we can use any frequency grid and Doppler shift in the frequency space 
is not just limited to neighboring bins but can happen in multiple bins between the two different sets of frequency grids $\nu_f$ and $\nu_{0,f}$ in the 
co-moving frame. Multi-frequency radiation moment equations can be also derived by integrating specific intensities over the angles \citep{MihalasKlein1982,Vaytetetal2011,Zhangetal2012,AnninosFragile2020}. These equations will need closure relations for the second and third moments of 
specific intensities in order to evolve the frequency dependent radiation energy density and flux.  In contrast, our equations evolving specific intensities are closed and no additional closure assumption is needed. 

\section{Numerical Algorithm}
\label{sec:num}
Our numerical scheme to solve this set of multi-group radiation MHD equations  is a natural extension of the implicit solver for grey RT as described in \cite{Jiang2021}. The transport term $c\bn\cdot \bfnabla I_f$ for different frequency groups is independent of each other, which is one big advantage of evolving specific intensities in the lab frame. This term is discretized in the same way as described in Section 3.2.1 of \cite{Jiang2021} and we will not repeat here.  Before we describe how the source terms in different frequency groups are coupled together as well as the Compton terms, we will first give an overview of the basic steps.                                                

\subsection{Steps of the Multi-Group Radiation MHD Scheme}
The numerical scheme described below is for the second order Van-Leer integrator with predict-correct steps in {\sf Athena++} \citep{Stoneetal2020}. Details for some of the steps will be described in the following sections. 

\begin{enumerate}[label=  Step \arabic{enumi}:,leftmargin=*,labelindent=1em]
 \item Set up the spatial mesh grid for both the fluid and radiation field. Choose angular and frequency grids for  specific intensities. Initialize both MHD and radiation variables. Calculate time step $\Delta t$ based on the standard Courant-Friedrichs-Lewy condition for MHD.
 \item Evolve the MHD equations for half time step $\Delta t/2$ using the standard MHD integrator in {\sf Athena++}.
 \item Construct the matrix coefficients for the RT equation with half time step $\Delta t/2$ based on fluid variables at the beginning of the step. The coefficients will include contributions from the transport term as well as the source term $S_f(I)$. 
 
 \item Solve the RT equation iteratively for specific intensities until a desired precision is achieved. 
 The change of specific intensities due to the Compton term $S_c(I)$ is also added in the co-moving frame during the iteration.
 \item Add radiation energy and momentum source terms to the gas based on total energy and momentum conservation to get the final gas properties in the predict step. 
 \item Repeat steps 2 to 5 to update both MHD and radiation variables for a full time step $\Delta t$.
\end{enumerate}

\subsection{Implicit Solver for Multi-group RT}
\label{sec:solve_source}

For any step $m$,  we start with the initial condition of gas and radiation field given by $\rho^m, T^m, \bv^m$ and $I_{f}^m$. Then we have the following equations to advance the radiation field to the next step $m+1$ by time step $\Delta t$ with the change of gas internal energy as
\begin{eqnarray}
	I_{f}^{m+1}-I_f^m+\Delta t c\bn\cdot \bfnabla I_f^{m+1}=\nonumber \\
		\Delta t c\Gamma^{-3} \mathcal{M}_{\nu}^{-1}\left[ \right. 
	\rho^{m}\left(\kappa_s+\kappa_{a,f}\right)\left(J_{0,f}^{m+1}-I_{0,f}^{m+1}\right)\nonumber\\
	+\rho^{m}\kappa_{p,f}\left(\varepsilon_{0,f}^{m+1} - J^{m+1}_{0,f}\right)  + S_c(I_f^{m+1})  \left.	\right],\nonumber\\
	\frac{\rho^m k_B}{\left(\gamma_g-1\right)\mu m_p}\left(T^{m+1}-T^m\right)=-\gamma 4\pi \Delta t c \nonumber\\
	\times \sum_{f=0}^{N_f-1} \left[\rho^m\kappa_{p,f}\left(\varepsilon_{0,f}^{m+1}-J_{0,f}^{m+1}\right)+S_c(I_f^{m+1})\right].
	\label{eq:discretized_equation}
\end{eqnarray}
Here we calculate the opacities $\kappa_s,\kappa_{a,f}$ and $\kappa_{p,f}$ using fluid quantities at time step $m$ based on analytical functions provided during the run time 
or bi-linear interpolation of provided opacity tables as a function of $\rho$ and $T$. We assume they do not change during this step.
The emissivity in each frequency group needs to be calculated with the advanced temperature $T^{m+1}$ for the fixed frequency grid $\nu_f$. 
For our default assumption of local thermal equilibrium, it will be 
\begin{eqnarray}
	\varepsilon_{0,f}^{m+1}=\frac{15a_r\left(T^{m+1}\right)^4}{4\pi^5}\times \nonumber\\
	\int_{h\nu_f/k_BT^{m+1}}^{h\nu_{f+1}/k_BT^{m+1}}\frac{\nu_T^3 d\nu_T}{\exp(\nu_T)-1}.
	\label{eq:emissivity}
\end{eqnarray}

The transport term $\bn\cdot \bfnabla I_f^{m+1}$ is calculated in the same way as done in \cite{Jiang2021}, which can be expressed as linear combinations of $I_f^{m+1}$ at neighboring spatial locations. But the source coupling terms are highly non-linear with respect to gas temperature $T^{m+1}$. This is a set of equations for $N_f\times N\times N_x\times N_y\times N_z$ specific intensities plus gas temperature, where $N_x,N_y,N_z$ are total number of cells along three spatial coordinates. The transport term couples specific intensities at different spatial locations together and the source terms couple specific intensities with different angles and frequencies together. Therefore all these equations need to be solved simultaneously via an iterative approach. 

For specific intensities at spatial location $(i,j,k)$, we iterate the following equations 
\begin{eqnarray}
	&&g_1 I_{f,l}^{m+1}+g_2 I_{f,l-1}^{m+1}(i-1)+g_3 I_{f,l-1}^{m+1}(i+1)\nonumber \\
	&&+g_4 I_{f,l-1}^{m+1}(j-1)+g_5 I_{f,l-1}^{m+1}(j+1) \nonumber \\
	&&+g_6 I_{f,l-1}^{m+1}(k-1)+g_7 I_{f,l-1}^{m+1}(k+1) \nonumber \\
	&=&I_f^m-\Delta t \bfnabla \cdot\left(\mathscr{F}\bv^m I_f^m\right) \nonumber \\
	&+&\Delta t c \Gamma^{-3}\mathcal{M}_{\nu}^{-1}\left[  \right.
 \rho^m\left(\kappa_s+\kappa_{a,f}\right)\left(J_{0,f,l}^{m+1}-I_{0,f,l}^{m+1}\right) \nonumber \\
 &+&\rho^m \kappa_{p,f}\left(\varepsilon_{0,f,l}^{m+1}	- J_{0,f,l}^{m+1} \right)+ S_c(I_{f,l}^{m+1}) \left.	\right],\nonumber\\
 &&\frac{\rho^m k_B}{\left(\gamma_g-1\right)\mu m_p}\left(T^{m+1}_l-T^m\right)=-\gamma 4\pi \Delta t \nonumber\\
 &&	\sum_{f=0}^{N_f-1}\left[ \rho^m \kappa_{p,f}\left(\varepsilon_{0,f,l}^{m+1}-J_{0,f,l}^{m+1}\right)+S_c\left(I_{f,l}^{m+1}\right)\right].
\end{eqnarray}
Here $l$ represents the step number during the iterative process and we take $I^{m+1}_{f,0}=I^m_f$. The coefficients $g_1, g_2, g_3, g_4, g_5, g_6, g_7$ and $\mathscr{F}$ are coming from discretization of the transport term for each frequency group and they are calculated in the same way as described in \cite{Jiang2021}, which we will not repeat here. The mean radiation energy density is simply the sum of all the specific intensities at the same step $J^{m+1}_{0,f,l}=\sum_{n=0}^{N-1} w_n^{\prime}I^{m+1}_{0,f,l}$. Notice that the angular quadrature weight in the co-moving frame $w_n^{\prime}$ is normalized so that $\sum_{n=0}^{N-1}w_n^{\prime}=1$. 
The iteration always take specific intensities in the neighboring zones from last iterative step and then solve $I^{m+1}_{f,l}$ for all angles at the same spatial location together. Once the process converges, we get the solution to equation \ref{eq:discretized_equation}. 

Let's define all the known specific intensities at the beginning of  step $l$ as
\begin{eqnarray}
	I_{f,c}&\equiv& I_f^m-\Delta t\bfnabla\cdot\left(\mathscr{F}\bv^m I_f^m\right) \nonumber\\
	&-&g_2 I_{f,l-1}^{m+1}(i-1) - g_3 I_{f,l-1}^{m+1}(i+1)\nonumber\\
	&-&g_4 I_{f,l-1}^{m+1}(i-1) - g_5 I_{f,l-1}^{m+1}(i+1)\nonumber\\	
	&-&g_6 I_{f,l-1}^{m+1}(i-1) - g_7 I_{f,l-1}^{m+1}(i+1).
\end{eqnarray}
After multiplying $\mathcal{M}_{\nu}\Gamma^4$ on both sides of the above equation, we get
\begin{eqnarray}
	g_1 I^{m+1}_{0,f,l}&=&\mathcal{M}_{\nu} \Gamma^4 I_{f,c} \nonumber\\
	&+&\Delta tc \Gamma \left[ \right.  \rho^m\left(\kappa_s+\kappa_{a,f}\right)\left(J_{0,f,l}^{m+1}-I_{0,f,l}^{m+1}\right) \nonumber\\
	&+&\rho^m \kappa_{p,f} \left(\varepsilon_{0,f,l}^{m+1} - J_{0,f,l}^{m+1}\right)	+S_c(I_{0,f,l}^{m+1}) \left.\right].
	\label{eq:solve_source}
\end{eqnarray}
Since the source term due to Compton process is too complicate to be solved with other terms together, we first solve this equation without $S_c(I_{0,f,l}^{m+1})$.  Then we can reorganize equation \ref{eq:solve_source} as
\begin{eqnarray}
	I_{0,f,l}^{m+1}=\left[ g_1+\Delta t c\Gamma\rho^m\left(\kappa_s+\kappa_{a,f}\right)\right]^{-1}\nonumber\\
	\left[ \right. \mathcal{M}_{\nu}  \Gamma^4I_{f,c} + \Delta t c\Gamma \rho^m\left(\kappa_s+\kappa_{a,f}\right)J_{0,f,l}^{m+1} \nonumber\\
   + \Delta t c\Gamma \rho^m\kappa_{p,f}\left(\varepsilon_{0,f,l}^{m+1}-J_{0,f,l}^{m+1}\right)	\left. \right].
   \label{eq:I_l_solution}
\end{eqnarray}
We sum over all the angles in the co-moving frame with weight $w_n^{\prime}$ and get
\begin{eqnarray}
J_{0,f,l}^{m+1}&=&\sum_{n=0}^{N-1}\frac{w_n^{\prime}\mathcal{M}_{\nu}\Gamma^4I_{f,c}}{g_1+\Delta t c \Gamma\rho^m\left(\kappa_s+\kappa_{a,f}\right)} \nonumber\\
&+&\left[\sum_{n=0}^{N-1}\frac{w_{n}^{\prime}\Gamma}{g_1+\Delta t c\Gamma\rho^m\left(\kappa_s+\kappa_{a,f}\right)} \right]\times \Delta t c\rho^m\nonumber\\
&\times&\left[\left(\kappa_s+\kappa_{a,f}-\kappa_{p,f}\right)J_{0,f,l}^{m+1}
+\kappa_{p,f}\varepsilon_{0,f,l}^{m+1}\right].
\end{eqnarray}
Notice that we have assumed the opacity and emissivity are independent of angles in the co-moving frame but they can depend on frequencies.  This can be reorganized to an equation for $J_{0,f,l}^{m+1}$ as
\begin{eqnarray}
	(1-A_{1,f})J_{0,f,l}^{m+1}=A_{2,f}\varepsilon_{0,f,l}^{m+1}+A_{3,f},
	\label{eq:J_l_solution}
\end{eqnarray}
where the coefficients are defined as
\begin{eqnarray}
	A_{1,f}&=&\left[\sum_{n=0}^{N-1}\frac{w_{n}^{\prime}\Gamma}{g_1+\Delta t c\Gamma\rho^m\left(\kappa_s+\kappa_{a,f}\right)} \right]\nonumber\\
	&\times&\Delta t c\rho^m \left(\kappa_s+\kappa_{a,f}-\kappa_{p,f}\right),\nonumber\\
	A_{2,f}&=&\left[\sum_{n=0}^{N-1}\frac{w_{n}^{\prime}\Gamma}{g_1+\Delta t c\Gamma\rho^m\left(\kappa_s+\kappa_{a,f}\right)} \right]\nonumber\\
	&\times&\Delta t c\rho^m \kappa_{p,f}\varepsilon_{0,f,l}^{m+1},\nonumber\\
	A_{3,f}&=&\sum_{n=0}^{N-1}\frac{w_n^{\prime}\mathcal{M}_{\nu}\Gamma^4I_{f,c}}{g_1+\Delta t c \Gamma\rho^m\left(\kappa_s+\kappa_{a,f}\right)}.
\end{eqnarray}
This can be used to get the equation for gas temperature as
\begin{eqnarray}
	\frac{\rho^m k_B}{\left(\gamma_g-1\right) \mu m_p}\left(T^{m+1}_l-T^m\right)=-\gamma 4\pi \Delta t c\rho^m\nonumber\\
	\times\sum_{f=0}^{N_f-1}\kappa_{p,f}\left[\frac{1-A_{1,f}-A_{2,f}}{1-A_{1,f}}\varepsilon_{0,f,l}^{m+1}-\frac{A_{3,f}}{1-A_{1,f}}\right].
	\label{eq:get_tgas}
\end{eqnarray}
For our default emissivity given by equation \ref{eq:emissivity}, $\varepsilon_{0,f,l}^{m+1}$ is a function of $T^{m+1}_l$, which can be determined by the above equation. For the case of one frequency bin (grey approximation), it is a fourth order polynomial for $T_l^{m+1}$ and we can find the root easily. For the general case, we need to solve the equation iteratively as $\overline{\varepsilon}_{0,f,l}\equiv \varepsilon_{0,f,l}^{m+1}/\left(T_l^{m+1}\right)^4$ is also a function of $T_l^{m+1}$. We first calculate $\overline{\varepsilon}_{0,f,l}$ for each frequency bin using $T_{l-1}^{m+1}$ and solve equation \ref{eq:get_tgas} to get an estimate for $T_l^{m+1}$. Then we use this estimated temperature to update $\overline{\varepsilon}_{0,f,l}$ and improve the solution for $T_l^{m+1}$. It typically takes a few iterations to get a solution to equation \ref{eq:get_tgas} with relative error smaller than $10^{-6}$. With this solution, we calculate $J_{0,f,l}^{m+1}$ according to equation \ref{eq:J_l_solution} and then get $I_{0,f,l}^{m+1}$ according to equation \ref{eq:I_l_solution}. After that, we update $I_{0,f,l}^{m+1}$ with the Compton source term 
$S_c$ if necessary and the updated solution is labeled as $I_{0,f,l}^{\mprime+1}$. The detailed process to solve the Kompaneets equation is 
described in Section \ref{sec:num_compton}.
Finally, we get the lab frame specific intensity as $I_{f,l}^{m+1}=\Gamma^{-4}\mathcal{M}_{\nu}^{-1} I_{0,f,l}^{\mprime+1}$. 
This finishes one iterative step and we continue this iteration for the whole simulation domain until the relative error $\Delta I$ is smaller than a preset criterion, where $\Delta I$ is defined as
\begin{eqnarray}
	\Delta I\equiv \sum_{f,n,i,j,k}|I^{m+1}_{f,l}-I^{m+1}_{f,l-1}|/\sum_{f,n,i,j,k}I^{m+1}_{f,l}.
\end{eqnarray}
Here the sum is done over the whole simulation domain and for all angles and frequency bins. We also support the stopping criterion that the maximum relative error $|I^{m+1}_{f,l}-I^{m+1}_{f,l-1}|/I^{m+1}_{f,l}$ 
is smaller than a preset value.

After we update all the lab frame specific intensities to time step $m+1$, we determine the radiation energy and momentum source terms that we need to deposit 
to the gas based on total energy and momentum conservation in the lab frame in 
the same way as described in section 3.2.3 of \cite{Jiang2021}. The source terms include contributions from all frequency bins. 
This will automatically capture all the velocity dependent source terms caused by Lorentz transformation. 

\subsection{Kompaneets Solver for Compton Scattering}
\label{sec:num_compton}

In this section, we will describe how the Kompaneets equation \ref{eq:compton} is solved to account for the change of specific intensities in the co-moving frame due to Compton scattering during each iterative step. The Kompaneets equation is solved with a conservative implicit approach following \cite{ChangCooper1970}. We use gas temperature $T^{m+1}_{l}$ from each iterative step and assume it is a constant while the Kompaneets equation is solved. The initial condition of this solver is 
$I^{m+1}_{0,f,l}$ and the associated $J^{m+1}_{0,f,l}$ as described in last section before the Compton source term $S_c$ is added. Without loss of generality for each iterative step $l$ and time step $m$, we will drop subscript $l$ and upper script $m+1$ for these quantities.  For simplicity, we will also use dimensionless frequency $\tnu\equiv h \nu/\left(k_BT_0\right)$ for a fiducial temperature unit $T_0$.

For frequency bins $0$ to $N_f-2$, we estimate the photon occupation number $n_{f+1/2}$ at the center of each frequency bin $[\tnu_f, \tnu_{f+1})$ according to equation \ref{eq:jton} with the monochromatic radiation energy density estimated as $J_{0,f}/(\tnu_{f+1}-\tnu_{f})$. 
For the last frequency bin $\tnu\in [\tnu_{N_f-1},\infty)$, we assume the photon occupation number follows the spectrum shape $n_{\tnu}\propto \exp(-\tnu/T)$ with the normalization fixed by the requirement that $\int_{\nu_{N_f-1}}^{\infty} J_{0,\nu}d\nu$ 
should match $J_{0,N_f-1}$ . 

In order to calculate the fluxes $F(n_{\tnu})$, we need to get the values of $n_{f}$ at the faces of each frequency bin, which are determined based on reconstruction of $n_{f-1/2}$ and $n_{f+1/2}$ as
\begin{eqnarray}
	n_{f}=\delta_{f}n_{f-1/2}+(1-\delta_{f})n_{f+1/2}.
\end{eqnarray}
The following steps adopt the procedure described by \cite{ChangCooper1970} to determine the coefficient $\delta_{f}$.
We first calculate the total photon numbers $N=\sum_{f=0}^{N_f-1} n_{f+1/2} \tnu_{f+1/2}^2 (\tnu_{f+1}-\tnu_{f})$. Notice that for the last frequency bin, we just take the integral with the assumed spectrum shape. The steady state solution is $\tilde{n}_{\tnu}=1/(\lambda\exp[\tnu/T]-1)$ and the coefficient $\lambda$ is determined by the following constrain
\begin{eqnarray}
	\int_0^{\infty} \frac{\tnu^2}{\lambda \exp[\tnu/T]-1}d\tnu=N.
\end{eqnarray}
We require that the condition $F(\tilde{n}_{\tnu})=0$ (equation \ref{eq:compt_fn}) is satisfied in the current frequency grid, which means
\begin{eqnarray}
	T\frac{\tilde{n}_{f+1/2}-\tilde{n}_{f-1/2}}{\tnu_{f+1/2}-\tnu_{f-1/2}}+\tilde{n}_{f}\left(\tilde{n}_{f}+1\right)=0.
	\label{eq:get_delta}
\end{eqnarray}
This will result in a second order polynomial equation for $\delta_f$, which only has one root between $0$ and $0.5$. 
Notice that the steady state solution is only used to determine the 
reconstruction coefficients $\delta_f$. Time dependent evolution of the spectrum is fully captured. 

We solve the Kompaneets equation implicitly to make sure the scheme is stable in the optically thick regime when the Compton thermalization timescale is shorter than the 
time step. We start with the photon occupation number $n_{f+1/2}^{\mprime}$ as determined by the partially updated co-moving frame $J_{0,f,l}^{m+1}$ as described in the last section. Then we update them with the following equation
\begin{eqnarray}
	\frac{T_e}{c\rho\kappa_{\rm es}}\tnu_{f+1/2}^2\left(\tnu_{f+1}-\tnu_{f-1}\right)\frac{n_{f+1/2}^{\mprime+1}-n_{f+1/2}^{\mprime}}{\gamma\Delta t}=\nonumber\\
	\tnu^4_{f+1}F\left(n^{\mprime+1}_{f+1}\right)-\tnu^4_{f}F\left(n^{\mprime+1}_{f}\right).
\end{eqnarray}
The flux is calculated at the boundary of frequency bins using updated photon occupation number $n_{f+1/2}^{\mprime+1}$ except the stimulated emission term, which is calculated using $n_{f+1/2}^{\mprime}$. This means
\begin{eqnarray}
	F\left(n^{\mprime+1}_{f}\right)&=&T\frac{n^{\mprime+1}_{{f+1/2}} - n^{\mprime+1}_{f-1/2}}{\tnu_{f+1/2}-\tnu_{f-1/2}}\nonumber \\
	&+&n^{\mprime+1}_{f}\left(1+n^{m}_{f}\right).
\end{eqnarray}
For the last frequency bin, we calculate the gradient using $n^{\mprime+1}_{N_f-3/2}$ and $n^{\mprime}_{N_f-1}$. However, if 
$n_{N_f-3/2}^{\mprime} < n^{\mprime}_{N_f-1}$, we simply set the flux at $\tnu_{N_f-1}$ to be 0. The flux at the frequency boundary $\tnu_0=0$ is also set to be 0. 
This results in a set of linear equations with respect to $n^{\mprime+1}_{f+1/2}$, which can be easily solved with Gauss elimination. 

Once we have the updated solution for $n_{f+1/2}^{\mprime+1}$, we can get the updated mean radiation energy density $J_{0,f,l}^{\mprime+1}$. Finally, the co-moving frame specific intensities are updated as
\begin{eqnarray}
	I_{0,f,l}^{\mprime+1}=I_{0,f,l}^{m+1}+(J_{0,f,l}^{\mprime+1} - J_{0,f,l}^{m+1}).
\end{eqnarray}

\section{Numerical Tests}
\label{sec:test}
The algorithm for multi-group radiation MHD developed here is designed to recover the grey approximation when one frequency bin is used, or if the conditions (initial condition, boundary condition and opacities) in all frequency bins are identical. We will not repeat all the tests that are already carried out in \cite{Jiang2021}. Instead, we will focus on testing the new multi-group features, including frequency dependent opacities and emissivities,  Doppler effects as well as the Kompaneets solver for Compton scattering. Since we will only be able to afford a small number of frequency groups for real applications, all the tests will use broad frequency groups instead of sharp features in the frequency space such as lines.  RT with opacities due to atomic lines can be also modeled in the framework developed here in principle. However, too many frequency groups will be needed to resolve the Doppler shift of lines to make it feasible in practice. We will discuss future extensions to model line transport more efficiently in section \ref{sec:discuss}. 

Results of the following tests will be reported in dimensionless numbers with assumed temperature, density, and length units to be $T_0$, $\rho_0$, and $L_0$, respectively. The velocity unit $v_0$ is the isothermal sound speed at temperature $T_0$ and time unit will be $L_0/v_0$. When we refer to a logarithmic frequency grid with $N_f$ frequency groups for $[\tnu_l, \tnu_r]$ in the following tests, we always mean two groups cover $[0,\tnu_l)$ and $[\tnu_r,\infty)$ respectively and the rest $N_f-2$ groups cover the range $[\tnu_l,\tnu_r)$ with logarithmic frequency spacing. Two dimensionless numbers that can convert the dimensionless numbers to physical units are $\Crat=c/v_0$ and $\Prat=a_rT_0^4\mu m_p/(\rho_0 k_BT_0)$ \citep{Jiangetal2012}, which we will report for each test problem. Gas adiabatic index is taken to be $\gamma_g=5/3$ unless otherwise specified. If not specified, we will assume local gas emission follows the blackbody spectrum (equation \ref{eq:emissivity0}).

\begin{figure}[htp]
	\centering
	\includegraphics[width=1.0\hsize]{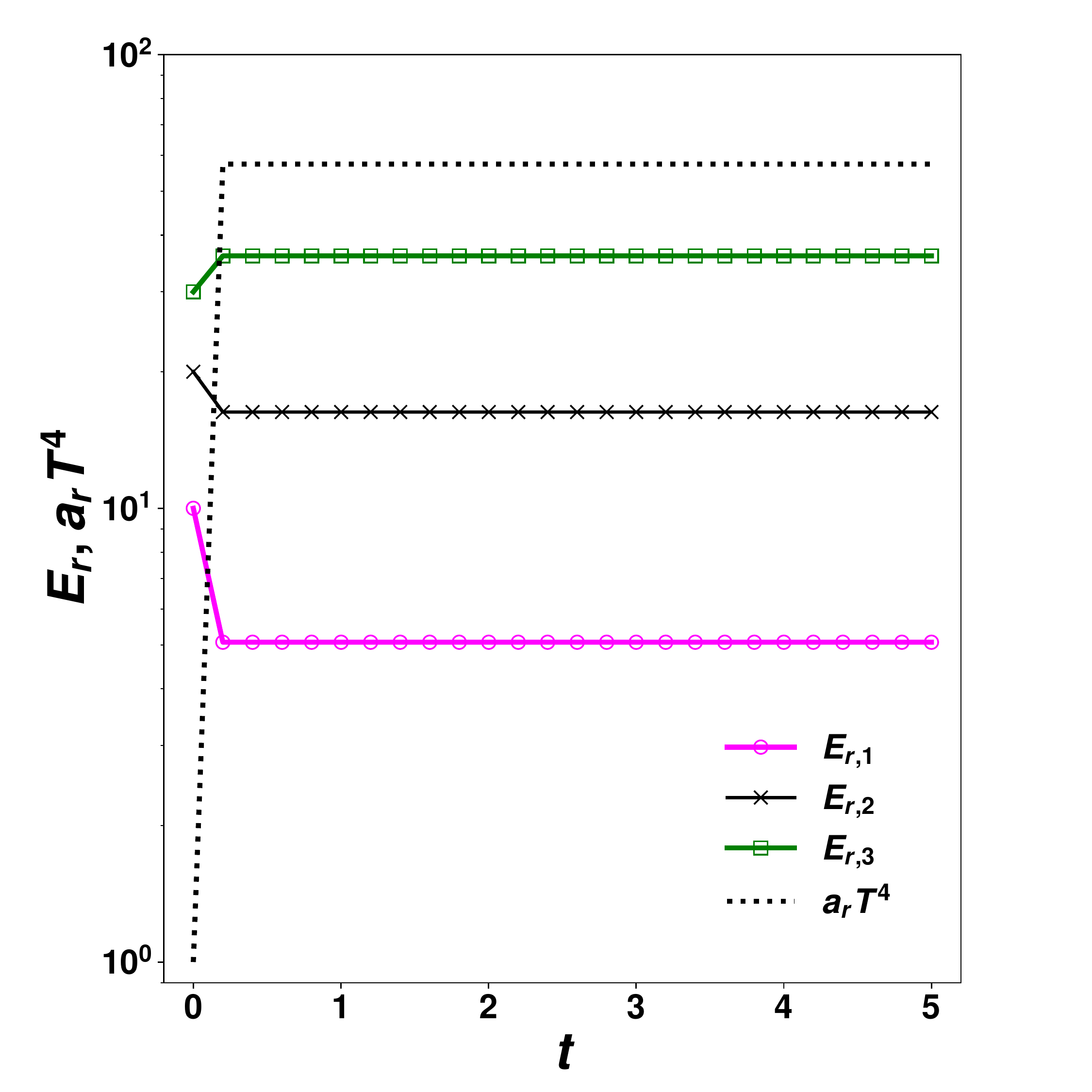}
	\caption{Histories of total gas emissivity $a_rT^4$ (the dotted black line) and radiation energy density in three frequency bins ($E_{r,0},E_{r,1},E_{r,2}$) for the thermal equilibrium test described in section \ref{sec:test_thermal_relaxation}. In steady state, radiation energy densities match the integrated blackbody emission in each frequency bin as determined using the steady state temperature. }
	\label{fig:thermal_relaxation}
\end{figure}

\subsection{Thermal Equilibrium of Multi-group Source Terms}
\label{sec:test_thermal_relaxation}
To test our solver for the source terms particularly when opacity varies significantly with frequencies, we create a 2D uniform box covering the region $[0,1]\times [0,1]$ with spatial resolution $32\times 32$. We use three frequency bins $\tnu\in [0,4),[4,8),[8,\infty)$ and the Planck mean absorption opacity in each 
bin is taken to be $\rho_0\kappa_pL_0=100,200,300$ respectively. The units are chosen so that $\Prat=1$ and $\Crat=10$. Density and temperature of the gas are set to be $1$ initially while the initial mean radiation energy density in the three frequency bins are $E_{r,0}/a_rT_0^4=10, E_{r,1}/a_rT_0^4=20$ and $E_{r,3}/a_rT_0^4=30$.  The thermalization timescales in all three frequency bins are smaller than the time step and solution reach steady state quickly as shown in Figure \ref{fig:thermal_relaxation}. Analytical solution in steady state can be calculated based on total energy conservation and the requirement that $4\pi \varepsilon_{0,f}$ needs to match $E_{r,f}$ in each frequency bin.  The gas temperature in steady state is $T=2.75$ and the emissivity in the three frequency bins are $4\pi \varepsilon_{0,0}/a_rT_0^4=5.07,4\pi \varepsilon_{0,1}/a_rT_0^4=16.30,4\pi \varepsilon_{0,2}/a_rT_0^4=36.00$, which agree with the radiation energy density in each bin as well as the analytical solution. Since frequency bins are fixed, the ratio $\varepsilon_{0,f}/a_rT^4$ also changes with $T$. In this test problem, it only takes five iterations of the polynomial solver  to get the updated gas temperature to the non-linear equation \ref{eq:get_tgas}. 

\begin{figure}[htp]
	\centering
	\includegraphics[width=1.0\hsize]{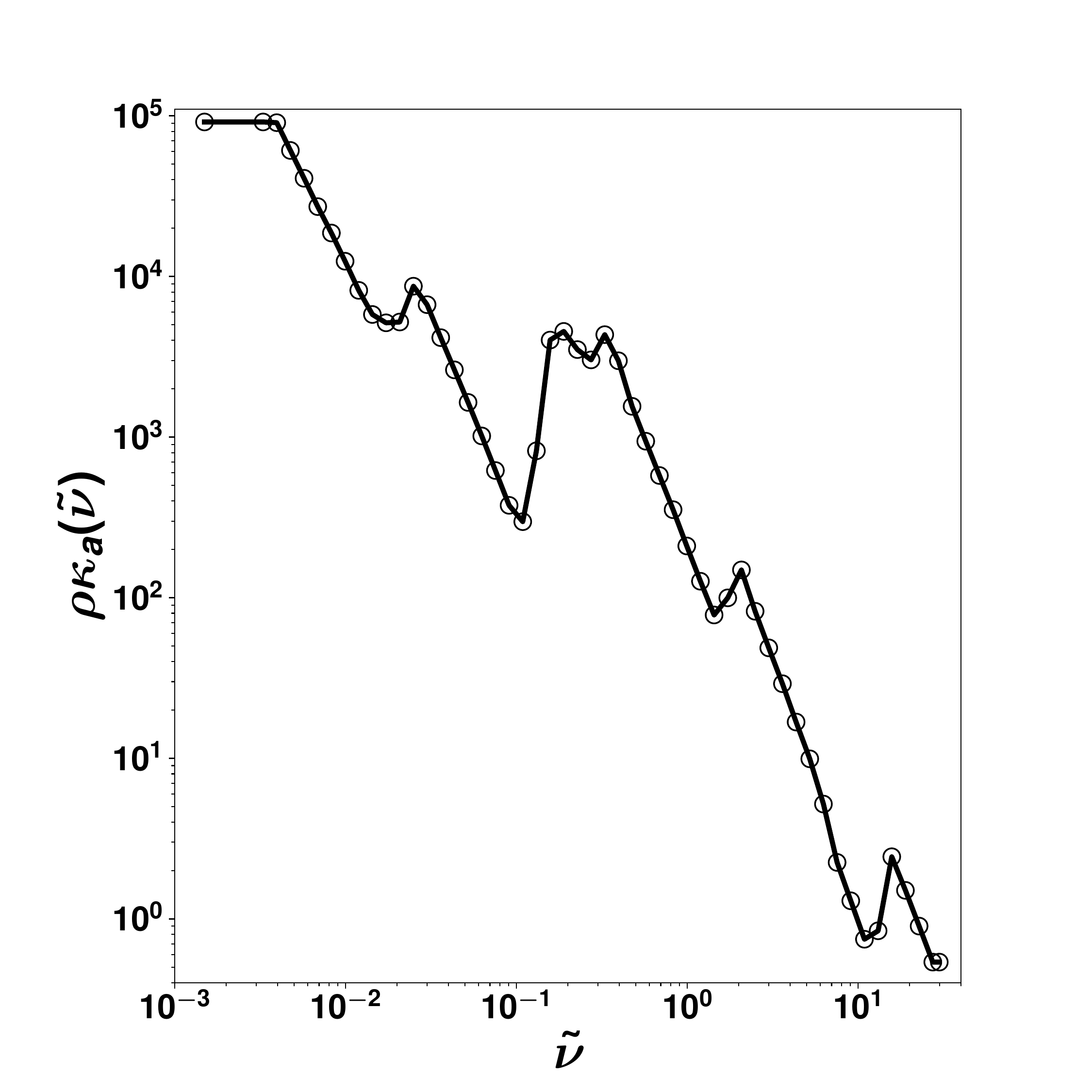}
	\includegraphics[width=1.0\hsize]{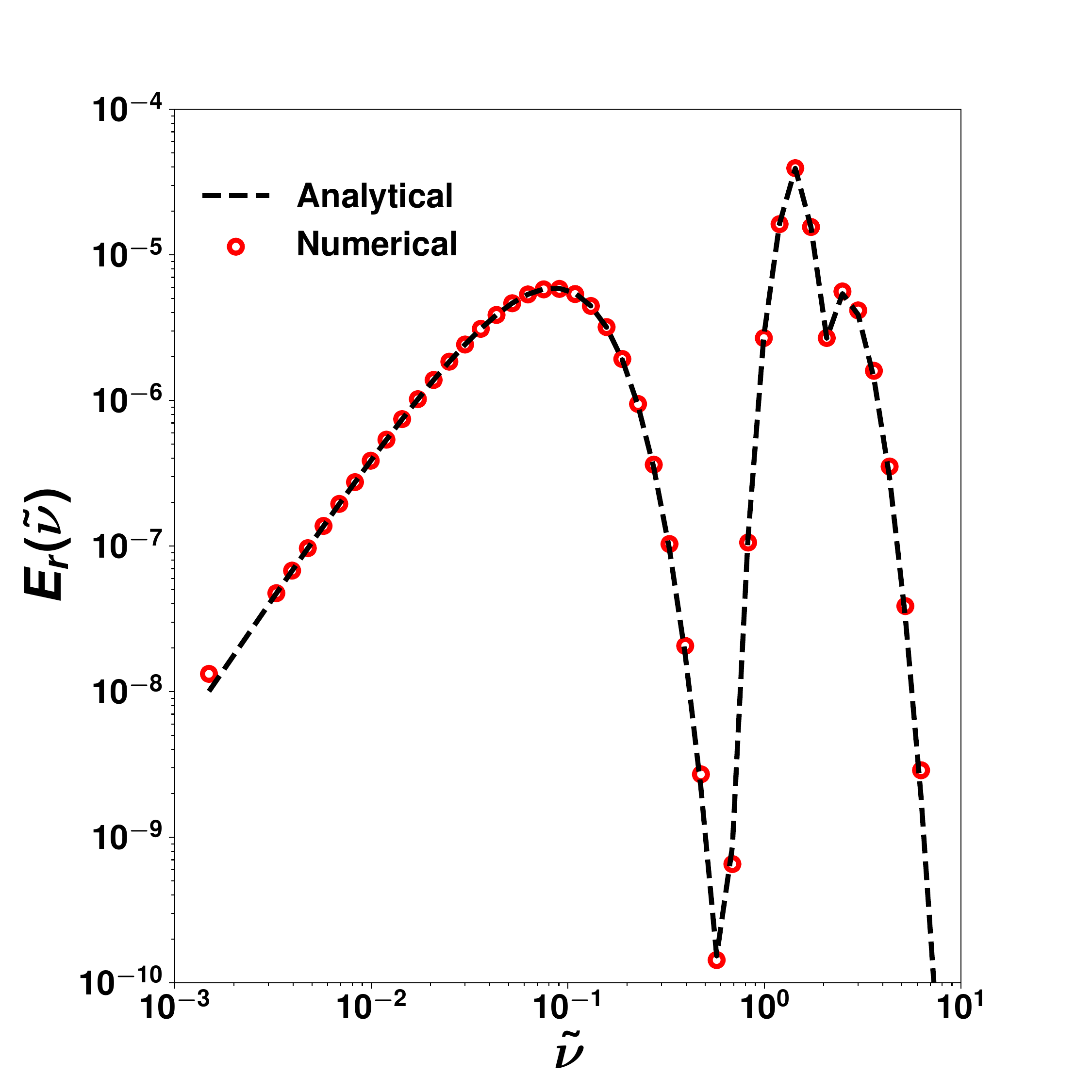}
	\caption{Top: absorption opacity coefficient used in the radiating sphere as described in section \ref{sec:radiating_sphere}. The frequency dependent opacity is taken from the tabulated value given by \cite{Graziani2008} for brominated plastic. The total optical depth across the simulation domain for each frequency $\tnu$ is simply $\rho\kappa_a$ times the radial box size 
		$0.03$.  
	Bottom: numerical solution (red dots) of the monochromatic radiation energy density in each frequency bin at radius $r=0.04$. This is the solution at  time $t=5.21\times 10^{-5}$. The dashed black line is the analytical solution using the opacity shown in the top panel.  Radiation energy density is scaled by $a_rT_0^4$ in this plot.}
	\label{fig:radiating_sphere}
\end{figure}

\subsection{Radiating Sphere}
\label{sec:radiating_sphere}
One useful test of the multi-group RT scheme is to compare the numerical and analytical solutions to the radiating sphere problem as proposed by \cite{Graziani2008}. This test is done in one dimensional spherical polar coordinate covering the radial range $r\in[0.02,0.05]$ with $32$ uniformly spaced grid points. Gas temperature is fixed to be $T_g=0.03$ with the temperature unit chosen so that the dimensionless speed of light $\Crat=805.9$. Gas velocity is 0 and it does not evolve in this test. At the inner boundary $r_i=0.02$, specific intensities are set to be isotropic with a blackbody spectrum at temperature $T_l=0.3$. At the outer boundary $r=0.05$, we simply copy all outgoing specific intensities from the last active zone to the ghost zones while set all incoming specific intensities to be 0. The absorption opacity coefficient $\rho\kappa_{a,f}$ is uniform in the whole simulation domain. Its dependence on frequency is tabulated by \cite{Graziani2008}, which is also shown in the top panel of Figure \ref{fig:radiating_sphere}. We use the second angular system as mentioned in section \ref{sec:angular_grid} with 40 angles covering $180^{\circ}$ with respect to the radial direction. Logarithmic frequency spacing is used to cover $[3\times 10^{-3},29.9]$ with $N_f=52$. Isotropic specific intensities are initialized in the simulation domain with blackbody spectrum at temperature $T_g$. Then radiation energy density at each radius $r$ and at any time $t$ is given by the following analytical solution \citep{Graziani2008}
\begin{eqnarray}
E_r(\tnu)&=&\left[B(\tnu,T_l)-B(\tnu,T_g)\right]f_{\tnu}(r,t) +B(\tnu,T_g),
\end{eqnarray}
where the function $f_{\tnu}(r,t)$ is defined based on the relative values of $\Crat t$ and $r-r_i$. The solution is basically the sum of local blackbody emission at temperature $T_g$ and the attenuated blackbody radiation from the 
inner boundary $r_i$ with temperature $T_l$ if the distance between the inner boundary and current location is smaller than $\Crat t$. Therefore, when $\Crat t<r-R$, $F_{\tnu}(r,t)=0$. For 
$r-r_i<\Crat t<\sqrt{r^2-r_i^2}$, we have
\begin{eqnarray}
f_{\tnu}(r,t)&=&\frac{r_i}{4r}\left\{ \left[1+\frac{r}{r_i}-\frac{1}{r_i\sigma_{\tnu}}\right]\exp\left(-\sigma_{\tnu}(r-r_i)\right) \right.\nonumber\\
&+& \left[\frac{1}{r_i\sigma_{\tnu}}-\frac{r_i}{\Crat t}\left(\frac{r^2}{r_i^2}-1\right)\right]\exp\left(-\sigma_{\tnu}\Crat t\right) \nonumber\\
&-& r_i\sigma_{\tnu}\left(\frac{r^2}{r_i^2}-1\right)\left[H(\sigma_{\tnu}(r-r_i)) \right. \nonumber\\
&-&\left. \left. H(\sigma_{\tnu}\Crat t)\right]\right\},
\end{eqnarray}
where we have defined the opacity coefficient $\sigma_{\tnu}\equiv \rho\kappa_a(\tnu)$ and the exponential integral
\begin{eqnarray}
	H(x)=\int_x^{\infty}\frac{\exp(-u)}{u}du.
\end{eqnarray}
When $\Crat t>\sqrt{r^2-r_i^2}$, the function $f_{\tnu}(r,t)$ is defined as
\begin{eqnarray}
f_{\tnu}(r,t)&=&\frac{r_i}{4r}\left\{ \left[1+\frac{r}{r_i}-\frac{1}{r_i\sigma_{\tnu}}\right]\exp\left(-\sigma_{\tnu}(r-r_i)\right) \right.\nonumber\\
&+& \left[\frac{1}{r_i\sigma_{\tnu}}-\sqrt{\frac{r^2}{r_i^2}-1}\right]\exp\left(-\sigma_{\tnu}\sqrt{r^2-r_i^2}\right) \nonumber\\
&-& r_i\sigma_{\tnu}\left(\frac{r^2}{r_i^2}-1\right)\left[H(\sigma_{\tnu}(r-r_i)) \right. \nonumber\\
&-&\left. \left. H\left(\sigma_{\tnu}\sqrt{r^2-r_i^2}\right)\right]\right\}.
\end{eqnarray}
Notice that this  is different from the commonly used solution under diffusion approximation \citep{Gonzalezetal2015}, as the high frequency bins are actually optically thin. 

To calculate the solution numerically, we take the Rosseland mean opacity for each frequency bin $\kappa_{a,f}$ to be the monochromatic value shown in Figure \ref{fig:radiating_sphere} at frequency $(\tnu_f+\tnu_{f+1})/2$. The numerical solution at radius $r=0.04$ and time $t=0.042/\Crat $ is shown as red dots in the bottom panel of Figure \ref{fig:radiating_sphere}. To get the monochromatic radiation energy density, we simply take $E_{r,f}/(\tnu_{f+1}-\tnu_{f})$. The numerical solution agrees very well with the analytical solution shown as the dashed line in the same figure. In the low frequency part ($\tnu<0.6$) where the domain is optically thick, the spectrum is dominated by local blackbody emission at temperature $T_g=0.03$. For the high frequency part ($\tnu > 0.6$), it is dominated by the attenuated emission from the inner boundary $r_i$. Notice that for the first bin, $E_{r,0}$ still agrees with the frequency integrated analytical solution for $\tnu\in[0,3\times 10^{-3}]$ very well. 
The estimated monochromatic radiation energy density in the first bin, which is not the quantity that we evolve, is slightly above the analytical value at $\tnu=1.5\times 10^{-3}$ because we assume the spectrum is constant inside that bin.

\begin{figure}[htp]
	\centering
	\includegraphics[width=1.0\hsize]{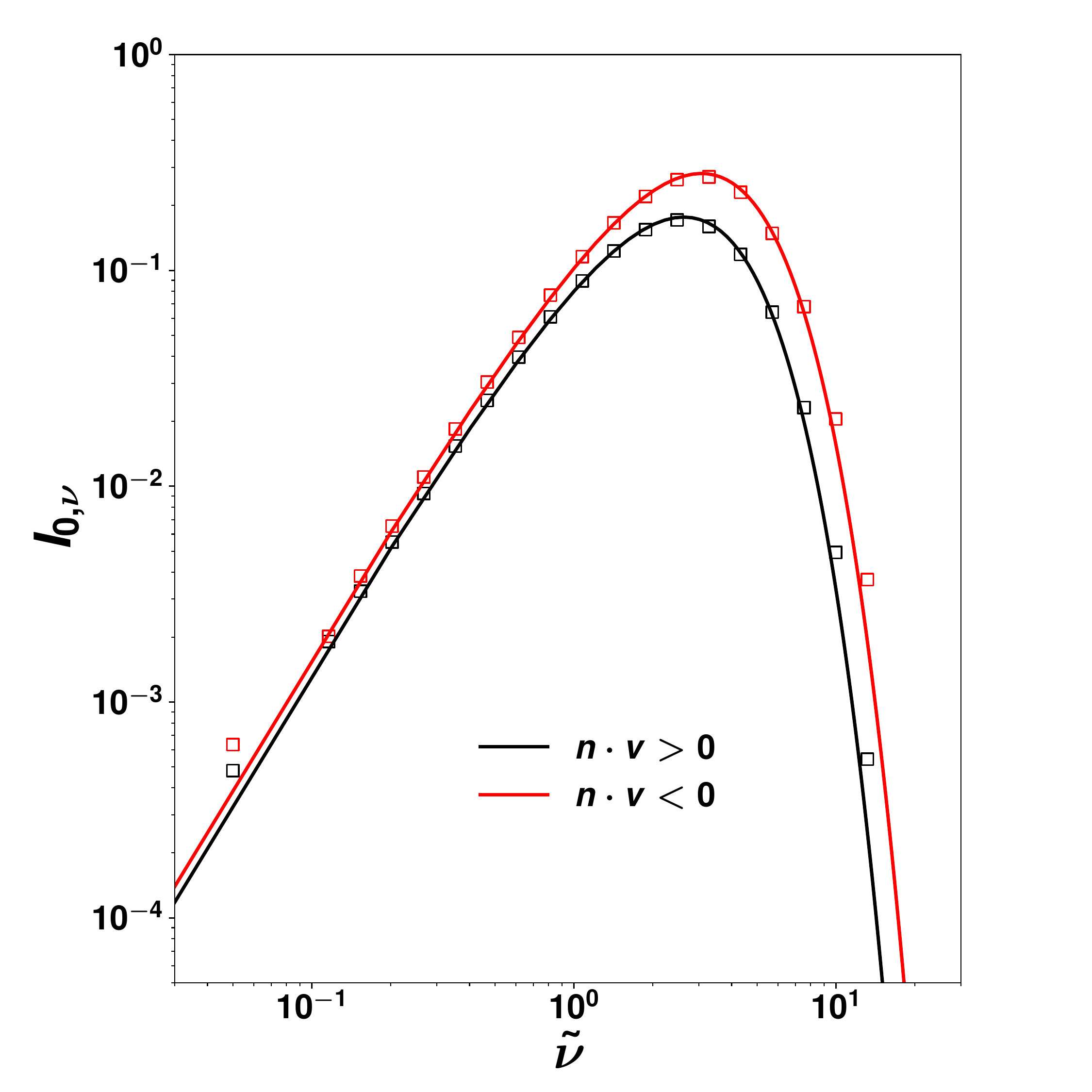}
	\caption{Spectra of co-moving frame monochromatic specific intensities with flow velocity $0.134c$ and isotropic blackbody spectrum in the lab frame. Black and red lines are for intensities propagating along two different directions  as indicated in the 
		figure. The black and red squares are numerically estimated co-moving frame specific intensities at the center of each frequency bin based on the frequency integrated intensities in each bin.  }
	\label{fig:doppler_shift}
\end{figure}

\subsection{Tests of Frame Transformation}
\label{sec:test_frame}
The frequency shift due to Doppler effect is handled by the Lorentz transformation as given by equation \ref{eq:I_transform} for specific intensities between lab and co-moving frames. We first test that the transformation is satisfied numerically and then show that it can correctly capture the frequency dependence of opacity. 

\begin{figure}[htp]
	\centering
	\includegraphics[width=1.0\hsize]{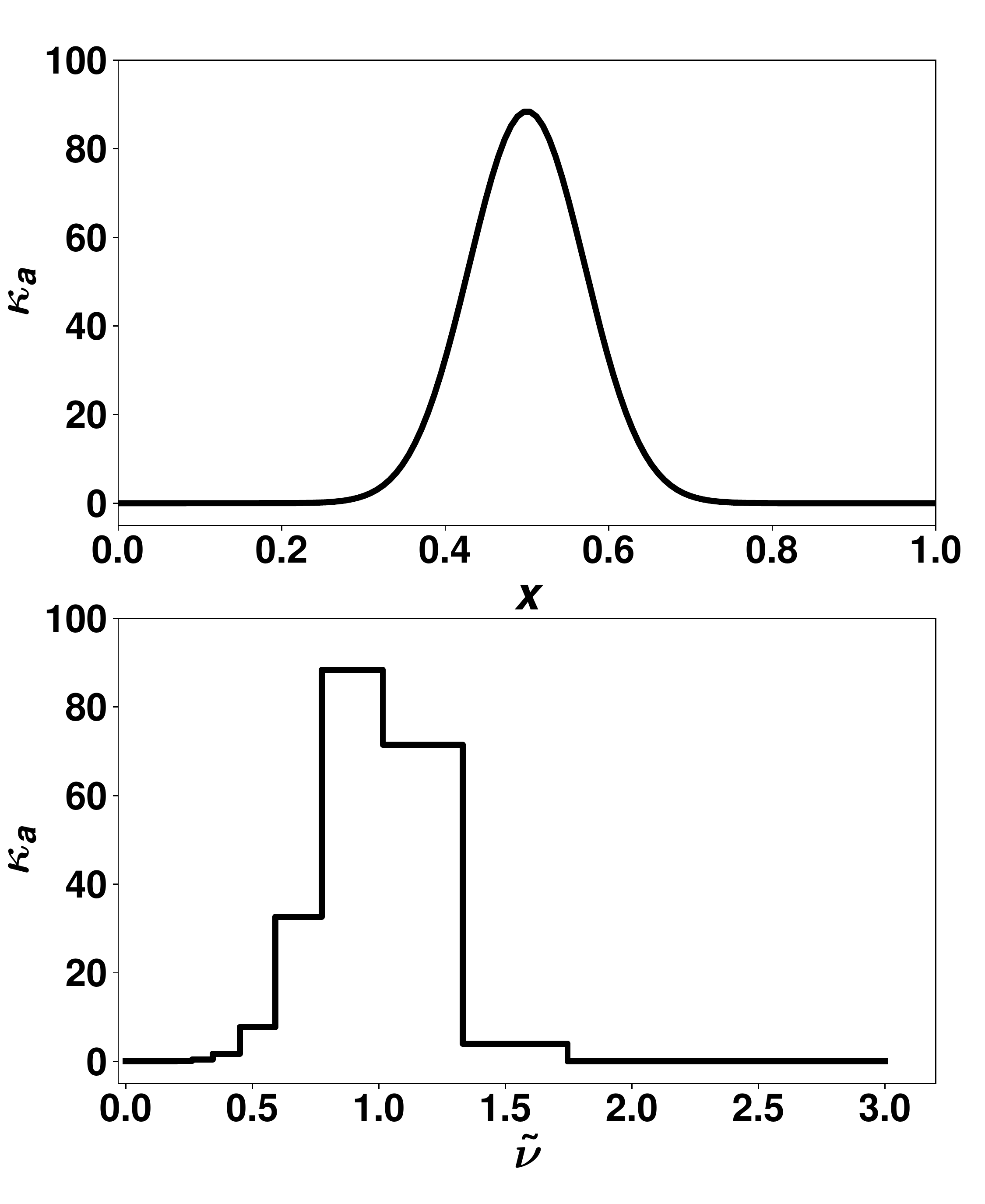}
	\caption{Opacity $\kappa_a$ used in the test of absorption due to a moving gas described in section \ref{sec:test_frame}. The top panel shows the spatial variation of $\kappa_a$ at $\tnu=0.9$ while the bottom panel shows the value of 
	$\kappa_a$ in each frequency bin at the location $x=0.496$.  }
	\label{fig:moving_absorption_opacity}
\end{figure}

We set up a 1D periodic domain covering the region $[0,1]$ with 16 grid points. We use 20 frequency groups and cover $\tnu\in[0.1,15]$ logarithmically. As a first step to just test the frame transformation, all the opacities are 0 so that there is no gas and radiation interaction.  We simply measure the radiation quantities in the lab and co-moving frames respectively. For simplicity, we only use two angles in each cell and initialize the specific intensities with a blackbody spectrum in the lab frame at temperature $T=1$. A uniform velocity $v=0.134 c$ is applied to the gas. The system is already in steady state and the frame transformation does not cause any change of the solution numerically by design. Theoretically expected spectra for the specific intensities along two different directions are shown as black and red lines in Figure \ref{fig:doppler_shift}. The intensity with $\bn\cdot\bv < 0$ gets boosted while the intensity with $\bn\cdot\bv > 0$ is reduced in the co-moving frame according to equation \ref{eq:I_transform}. Since we only evolve the frequency integrated intensities in each bin, we can estimate the monochromatic values at the center of each frequency bin by $I_{0,f}/(\nu_{f+1}-\nu_f)$. This can be done for the first 19 bins, which are shown as black and red squares in Figure \ref{fig:doppler_shift}. 
Our numerical solution $I_{0,f}$ always match the theoretically expected frequency integrated values in each bin. Between $\tilde{\nu}=0.1$ and $\tilde{\nu}=15$ where the frequency space is resolved with 18 bins, the monochromatic intensities also agree with the analytical values. Our numerical estimate at $\tilde{\nu}=0.05$ is always larger because the spectrum is assumed to be flat in each bin. 

\begin{figure}[htp]
	\centering
	\includegraphics[width=1.0\hsize]{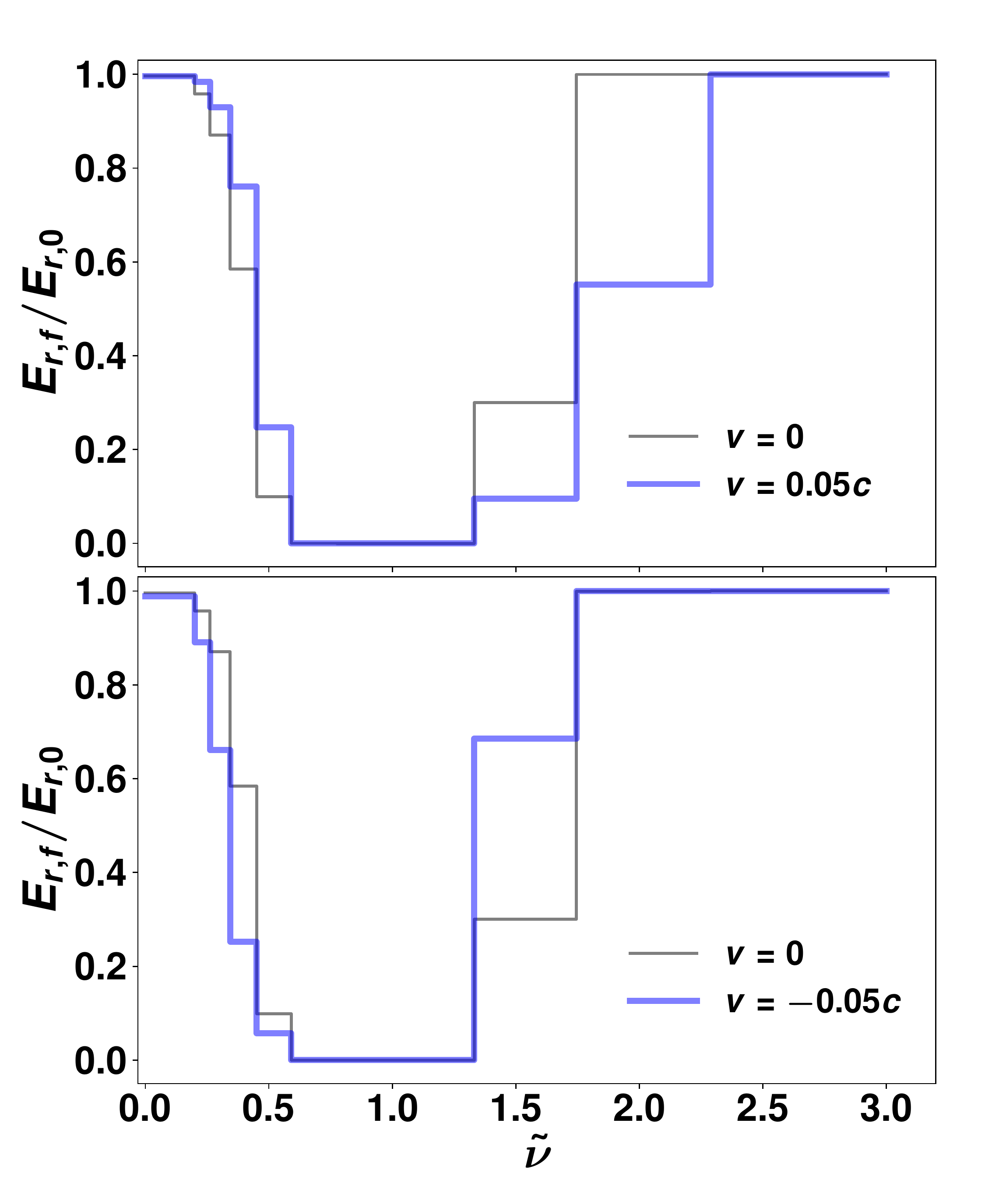}
	\caption{Spectra of the radiation energy density $E_{r,f}$ scaled by the injected value $E_{r,0}$ after the photons pass through the gas with opacity shown in Figure \ref{fig:moving_absorption_opacity} as described in section \ref{sec:test_frame}. The black lines show the results when gas is static, while the blue lines in the top and bottom panels show the results when gas has velocity $v=0.05c$ and $v=-0.05c$ respectively. }
	\label{fig:moving_absorption_hist}
\end{figure}

To test how the Lorentz transformation interacts with the source terms together, we use a frequency and spatial dependent absorption opacity defined in the co-moving frame as 
\begin{eqnarray}
	\kappa_{a,f}=\kappa_{p,f}&=&100\exp\left[-\left[((\tnu_f+\tnu_{f+1})/2-1)/0.3\right]^2\right] \nonumber\\
	&\times& \exp\left[-100\left(x-0.5\right)^2\right].
	\label{eq:moving_absorption_opacity}
\end{eqnarray}
The opacity in the last frequency bin is set to 0. Example profiles of the opacity as a function of spatial location and frequency are shown in Figure \ref{fig:moving_absorption_opacity}. Scattering opacity is 0 in this test. Gas density and temperature are fixed to be 1 for the whole simulation domain in this test and they do not evolve. Gas velocity is 0 in the first step. Total optical depth across the box reaches a maximum value $15.7$ in the frequency bin $\tnu\in[0.77,1.01)$ and then drops quickly for other bins. For $\tnu < 0.34$ or $\tnu > 1.33$, total optical depth across the box is smaller than $1$.  We still use two angles per cell for specific intensities in this test for simplicity and they are initialized to be 0. At the left boundary $x=0$,  the incoming specific intensities are set to a constant value $10^4 a_rT^4_0/4\pi$ for all frequency bins while outgoing specific intensities are set to 0. For the right boundary at $x=1$, we copy the outgoing specific intensities from the last active zone to the ghost zones while set incoming specific intensities to be 0. The injected radiation field will propagate through the simulation domain and get attenuated around $x=0.5$. We measure the steady state radiation energy densities in the lab frame as a function of frequencies at the right boundary, which are shown as the black lines labeled with $v=0$ in Figure \ref{fig:moving_absorption_hist}. The spectrum shape of $E_{r,f}$ basically follows the frequency dependence of opacity $\kappa_a$ as shown in Figure \ref{fig:moving_absorption_opacity}. For the two bins around $\tnu=1$ where the total optical depth is much larger than 1, $E_{r,f}$ approaches the local emissivity $\varepsilon_{0,f}$, which is taken to be blackbody emission with local gas temperature $T=1$. For other bins, the reduction radiation energy density follows the expected attenuation $\exp(-\tau_f)$, where $\tau_f$ is the integrated optical depth across the whole simulation domain for each frequency bin. 

We now keep everything the same but set the gas velocity to be $v=0.05c$. Since co-moving frame frequencies are related to the lab frame frequencies via equation \ref{eq:fre_transform} and the opacity given in equation \ref{eq:moving_absorption_opacity} is always applied in the co-moving frame, intensities with $\bn\cdot \bv>0$ will shift the attenuation to a lightly higher frequency and intensities with $\bn\cdot\bv < 0$ will move to the opposite direction. Because $E_{r,f}$ is dominated by the right propagating intensities, the spectrum of $E_{r,f}$ is also shifted to higher frequencies as shown in the top panel of Figure \ref{fig:moving_absorption_hist}. Similarly when gas has  velocity $v=-0.05c$, the spectrum of $E_{r,f}$ is shifted to lower frequencies as shown in the bottom panel of Figure \ref{fig:moving_absorption_hist}. The results can be quantitatively checked with the mixed frame formula to the first order of $v/c$  \citep{MihalasKlein1982}. If we neglect contribution from emissitivity, lab frame specific intensities are attenuated with lab frame monochromatic opacity $\kappa(\nu)\approx \kappa_0(\nu)\left[1-\left(\bn\cdot\bv/c\right)\left(1+\partial \log\kappa_0/\partial \log \nu\right)\right]$, where $\kappa_0(\nu)$ is the opacity given by equation \ref{eq:moving_absorption_opacity} but evaluated at the lab frame frequency $\nu$.  This is derived using the relation $\kappa(\nu)=\gamma\left(1-\bn\cdot\bv/c\right)\kappa_0(\nu_0)$. We average $\kappa(\nu)$ for each frequency bin to get the lab frame $\kappa_{a,f}$, which is then used to calculate the new attenuation $\exp(-\tau_f)$. This can explain the numerical solution very well except the bin $\tnu\in[0.6,1.3]$, where $E_{r,f}$ is determined by local emissivity as it is optically thick and the above simple analysis does not apply.

\subsection{Tests of the Kompaneets Solver}
In this section, we describe several tests of the Kompaneets solver implemented with our multi-group RT algorithm. Particularly, we will quantify the accuracy of the solver as a function of the number of frequency groups. 

\subsubsection{Spectrum Evolution with A Delta function Initial Condition}
\label{sec:compton_delta}

\begin{figure}[htp]
	\centering
	\includegraphics[width=1.0\hsize]{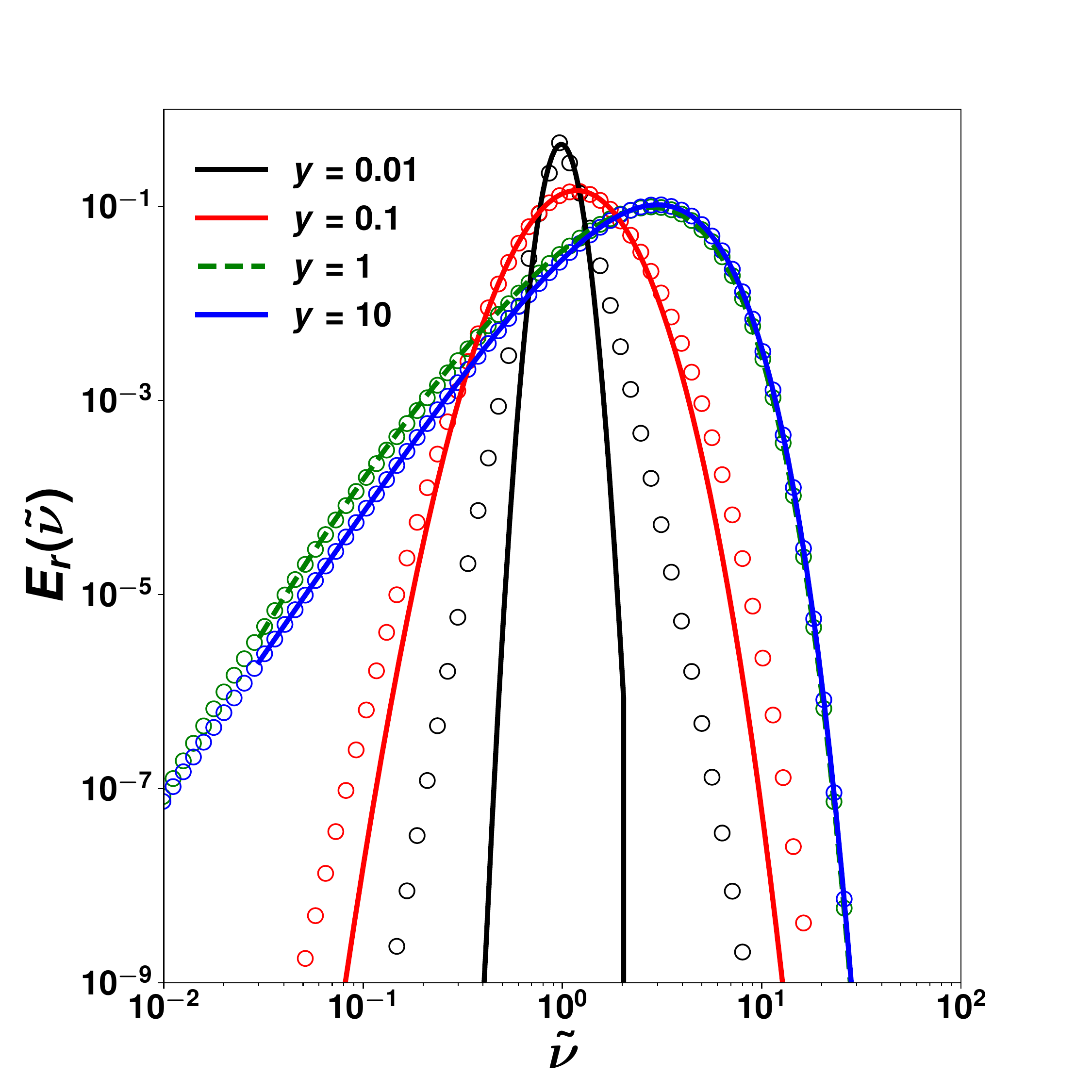}
	\caption{Comparing photon spectra due to Compton scattering without stimulated emission at different times for an initial delta function spectrum at $\tnu=1$ (section \ref{sec:compton_delta}). The circles are our numerical solutions for different $y$ values while the solid lines are the corresponding analytic solutions given by \cite{Becker2003}.   }
	\label{fig:compton_delta}
\end{figure}

If we neglect the stimulated emission term, Kompaneets equation becomes linear with respect to $n_{\tnu}$ and it can be solved analytically. \cite{Becker2003} gave an exact solution to describe the spectrum evolution at any time when the initial condition is the Green's function at a particular frequency. In order to compare our numerical solutions to 
the analytical solutions, we set the stimulated emission term to be 0 in the solver while keeping everything else the same. Time evolution of the solution is typically described by the parameter $y\equiv \left[c\rho\kappa_{\rm es} kT /(m_e c^2)\right] t$ and all gas quantities are fixed in this test. We choose the parameters and unit system such that $m_ec^2/kT=100$ and $c\rho\kappa_{\rm es}=1000$. The spatial and angular grids are not relevant for this test. We simply use a 1D domain from 0 to 1 with 32 grid points and two angles per cell. We use logarithmic frequency spacing to cover $\tnu\in [0.001,100]$ with $N_f=100$. We limit the time step to be $\Delta t=1.55\times 10^{-4}$ so that the thermalization process can be properly resolved as it is equivalent to $\Delta y=1.55\times 10^{-3}$. To mimic the delta function in the initial condition of the analytical solution, we initialize the frequency integrated specific intensities in the frequency bin $\tnu\in[0.91,1.02)$ to be 1 and 0 for all the others. The frequency dependent photon occupation number normalized by the number density of the initial delta function at frequency $\tnu_0$ for any $y$ parameter is (equation 33 of \citealt{Becker2003})
\begin{eqnarray}
&n_G&(\tnu,\tnu_0,y)=\frac{32}{\pi}e^{-9y/4}\tnu_0^{-2}\tnu^{-2}e^{(\tnu_0-\tnu)/2} \nonumber\\
&\times& \int_0^{\infty}e^{-u^2y}\frac{u\sinh(\pi u)}{(1+4u^2)(9+4u^2)}W_{2,iu}(\tnu_0)W_{2,iu}(\tnu)du \nonumber\\
&+&\frac{e^{-\tnu}}{2}+\frac{e^{-\tnu-2y}}{2}\frac{(2-\tnu)(2-\tnu_0)}{\tnu_0\tnu},
\end{eqnarray}
where $W_{2,iu}(\tnu)$ is the Whittaker function. This can be converted to the monochromatic radiation energy density $E_r(\tnu)$, which are shown as the solid lines in Figure \ref{fig:compton_delta} for different $y$ parameters with $\tnu_0=1$. For the numerical solution, we calculate $E_r(\tnu)$ as $E_{r,f}/(\tnu_{f+1}-\tnu_f)$, which are shown as open circles in the same figure. The numerical solution is more diffusive at the early time $y\leq 0.1$ but agrees with the analytical solution very well at the later time. The discrepancy at the earlier time is likely due to the fact that our initial condition is not a true delta function. This is also seen in other codes that performed similar test \citep{Narayanetal2016}.

\begin{figure}[htp]
	\centering
	\includegraphics[width=1.0\hsize]{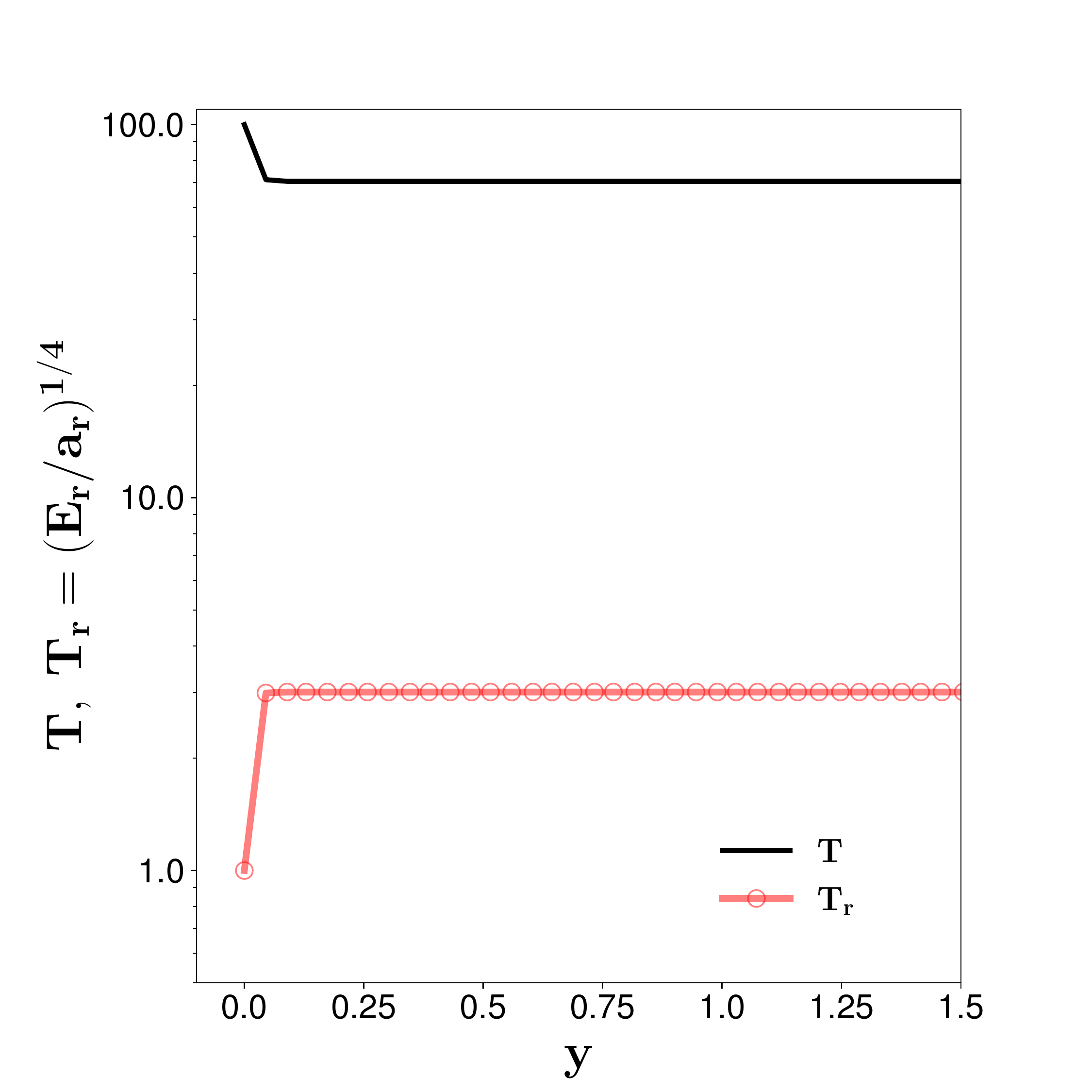}
	\includegraphics[width=1.0\hsize]{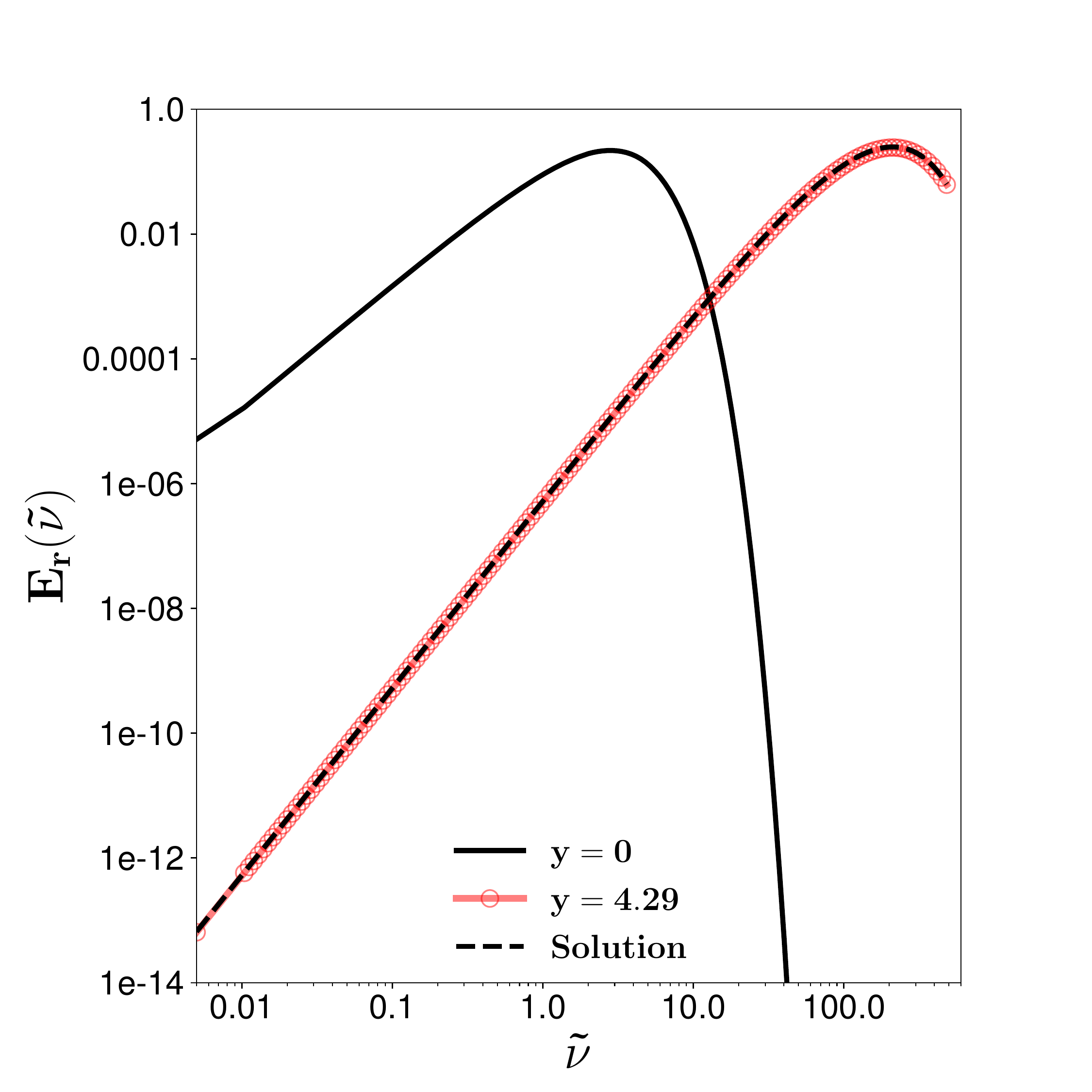}
	\caption{Thermalization of gas and radiation due to Compton scattering for the test described in section \ref{sec:compton_thermal}. The top panel shows 
		the history of gas temperature $T$ and effective radiation temperature $\left(E_r/a_r\right)^{1/4}$. The bottom panel shows spectra of the monochromatic radiation energy density 
		at $y=0$ and $y=4.29$. The dashed line is the analytic solution.  }
	\label{fig:compton_thermal}
\end{figure}

\begin{figure}[htp]
	\centering
	\includegraphics[width=1.0\hsize]{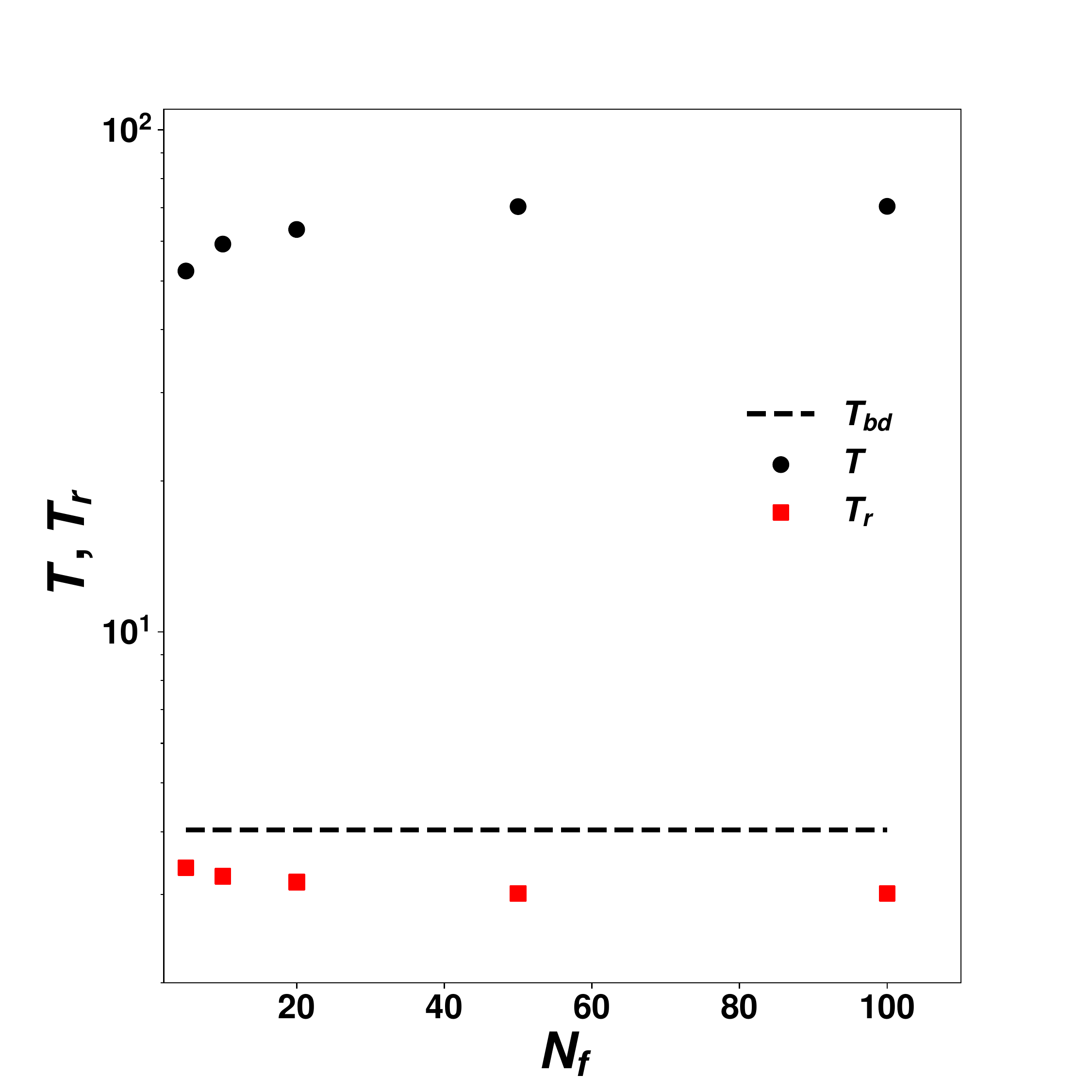}
	\caption{Gas and effective radiation temperature in steady state as a function of number of frequency bins $N_f$ used in the Compton thermalization test described in section \ref{sec:compton_thermal}. If we assume the radiation spectrum is black body without solving the Kompaneets equation,  the gas and radiation temperature will be the same in steady state, which is showed as the dashed black line $T_{bd}$. }
	\label{fig:compton_resolution}
\end{figure}

\subsubsection{Compton Thermalization}
\label{sec:compton_thermal}

Compton can be the dominant thermalization process for low density, high temperature gas such as the corona region of accretion disks. To test the capability of our Kompaneets solver to evolve the gas temperature 
in this scattering dominated regime, we setup a  1D domain covering $x\in[0,1]$ with 32 grid points and we use 2 angles per cell for specific intensities. 
Temperature unit is chosen to be $T_0=10^4$ K and initial gas temperature is $T=10^6$ K. Uniform density $\rho=10^{-10}$ g/cm$^3$ is adopted for the whole box. The opacity and time unit are chosen so that $y=\left[c\rho\kappa_{\rm es} kT /(m_e c^2)\right] t=428.99 t$. We use $N_f=150$ for the frequency grid and logarithmic spacing to cover $\tnu\in [0.01,500]$. We initialize the radiation field with a blackbody spectrum at temperature $T_0$ and let the system evolve to steady sate. Histories of gas temperature and effective radiation temperature $T_r\equiv (\sum_{f=0}^{N_f-1} E_{r,f}/a_r)^{1/4}$ are shown in the top panel of Figure \ref{fig:compton_thermal} and steady state spectrum is shown in the bottom panel of the same figure. The analytic solution can be calculated based on total energy conservation as well as the fact that total photon 
numbers remain the same for pure scattering. Photon occupation number in steady state should be $n_{\tnu}=1/\left[\lambda \exp(\tnu/T)-1\right]$, where the coefficient $\lambda$ is constrained by the total photon numbers in the initial condition. With another constraint of total energy conservation, we can determine the gas temperature in steady state, which is $T=70.46T_0$ and agrees with the numerical solution very well. This is also consistent with the Compton temperature as given by equation \ref{eq:compton_tem}. If we do not solve the Kompaneets equation but assume  photons have a black body spectrum, gas and radiation temperature needs to be the same in steady state, which will be $T_{bd}=4.03T_0$ in this case. Clearly, gas temperature will be significantly underestimated.

In real multi-dimensional simulations, it can be too expensive to use $150$ frequency bins coupled with the angular grid. It is interesting to see how the results will depend on our 
resolution in frequency space. The steady state gas and radiation temperature as a function of total number of frequency bins $N_f$ are shown in Figure \ref{fig:compton_resolution}. If only $5$ frequency bins are used, we get $T=52.4$. The error is reduced to $10\%$ when the frequency resolution is increased to $20$.  The solution already reaches the accurate value with $50$ frequency bins. It is interesting to see that even solving the Kompaneets equation using 5 frequency bins, the solution we get is much better compared with the case if blackbody spectrum is assumed.

\begin{figure}[htp]
	\centering
	\includegraphics[width=1.0\hsize]{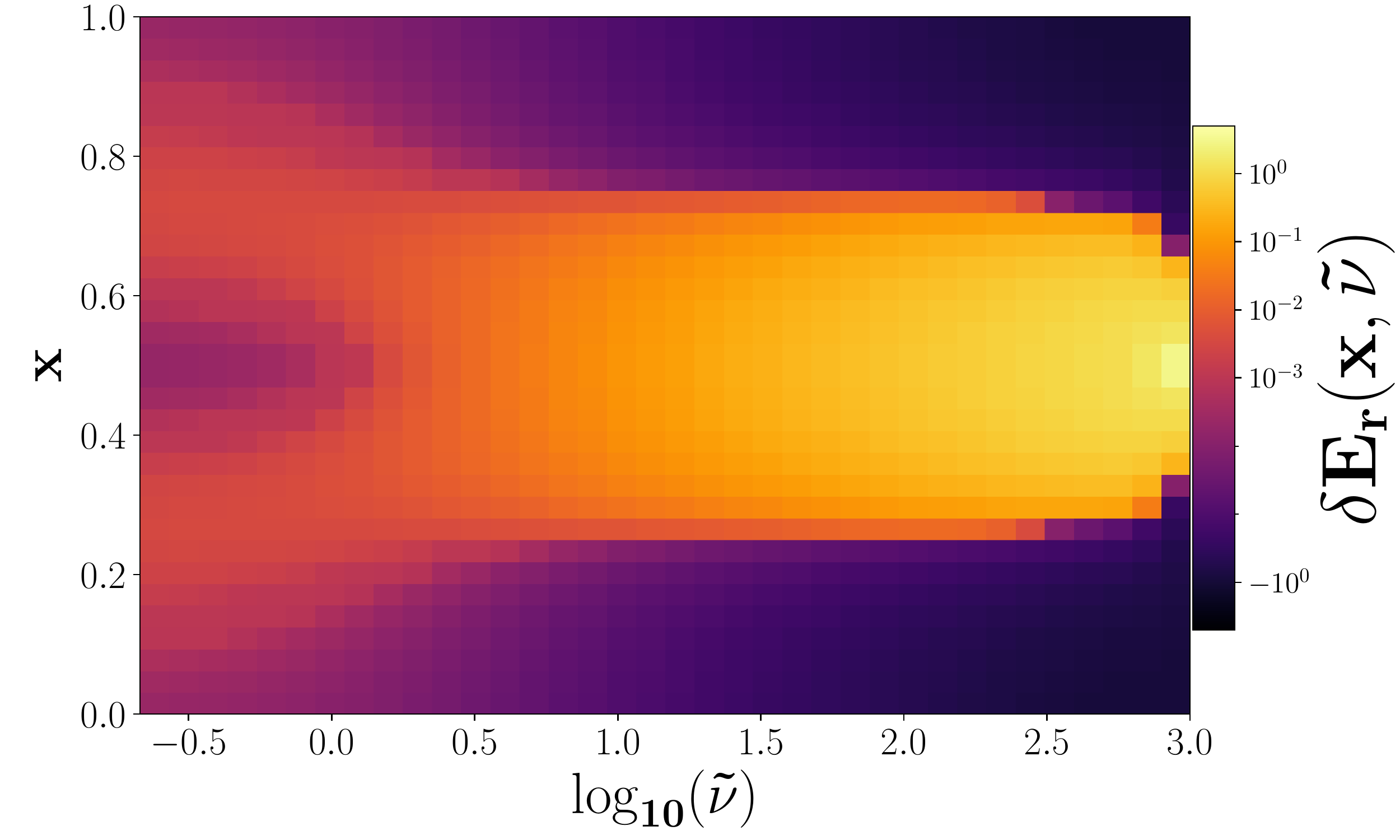}
	\caption{Spatial and frequency distribution of the monochromatic radiation energy density variations $\delta E_r(x,\tnu)$ (equation \ref{eq:delta_Er_x_nu}) at time $t=2$ for the bulk Compton test with a convergent/divergent flow described in section \ref{sec:bulk_compton}. The variation $\delta E_r(x,\tnu)$ represents relative fluctuations of $E_r(x,\tnu)$ scaled by the spatially averaged values for each frequency $\tnu$. The colormap is symmetric logarithmic, which means positive and negative values of $\delta E_r(x,\tnu)$ are shown separately in the logarithmic scale of $|\delta E_r(x,\tnu)|$. }
	\label{fig:bulk_compton_spatial}
\end{figure}

\subsubsection{Bulk Componization}

\label{sec:bulk_compton}

Gas mechanical energy can also be exchanged with radiation energy directly via the forms of radiation work (for convergent and divergent flows) and radiation viscosity (for shear flows) terms. When the bulk velocity of electrons $|\bv|$ is larger than $\sqrt{3k_BT/m_e}$ \citep{Psaltisetal1997}, Compton process by the bulk motions can be more important than Compton due to thermal motions of electrons. This can happen in strongly radiation pressure dominated flows around supermassive black holes, where bulk velocity of the flow or turbulent velocity can exceed thermal electron velocity \citep{PayneBlandford1981,Titarchuketal1997,Turollaetal2002,Socratesetal2004,KaufmanBlaes2016,Kaufmanetal2018,Zrakeetal2019}. In our multi-group algorithm developed here, scattering, absorption and emission are all calculated in the fluid rest frame. Bulk velocity of electrons relative to photons in different frequency group and different angular directions are taken care of by self-consistent Lorentz transformation.

\begin{figure}[htp]
	\centering
	\includegraphics[width=1.0\hsize]{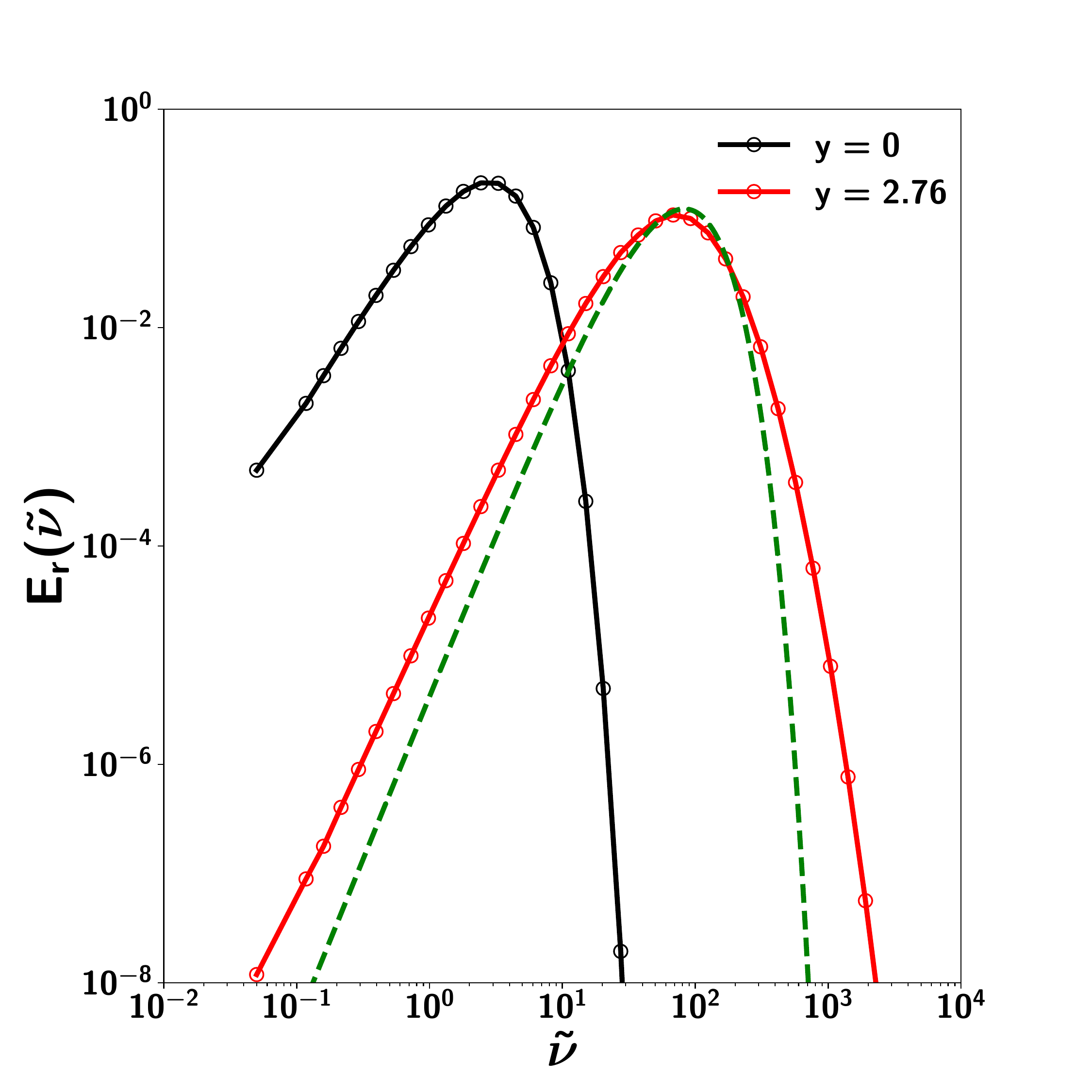}
	\caption{Spectrum of the spatially averaged monochromatic radiation energy density $E_r(\tnu)$ for the same test as shown in Figure \ref{fig:bulk_compton_spatial}. The black and red lines are for the initial blackbody spectrum and spectrum at time $t=2$ corresponding to the Compton $y$ parameter $2.76$. The dashed green line is a spectrum with Bose-Einstein distribution at the effective temperature $T+m_e v_m^2/(3k_B)=27.57T$, where $v_m=300$ is the amplitude of the sinusoidal velocity used in this test and gas temperature $T=1$ in our unit system. Chemical potential of the distribution is chosen so that total radiation energy density matches the numerical solution at this time. }
	\label{fig:bulk_compton_converge}
\end{figure}

To demonstrate how the bulk Compton effects are captured, we consider two simplified test problems for pure convergent/divergent flow and shear flow respectively. We first set up a 1D domain covering $x\in[0,1]$ with $32$ grid points. Gas temperature is fixed to $T_0=10^5$ K and the dimensionless speed of light $\Crat=8193.11$. Density unit is chosen so that the dimensionless radiation pressure unit $\Prat=1$. Constant density and scattering opacity are used so that 
$\rho\kappa_{\rm es}=10$, which means total scattering optical depth is 10 across the whole box. Absorption opacity is 0. We use 16 angles to resolve the angular distribution of specific intensities and logarithmic frequency spacing to cover $\tnu\in [0.1,10^4]$ with $N_f=40$.  Periodic boundary condition is used for the radiation field while gas quantities are not evolved in this test. Isotropic specific intensities are initialized with blackbody spectrum at temperature $T_0$. We first set the flow velocity to 0 and let the system evolve. We do confirm that our Compton solver is able to maintain the blackbody spectrum. Then we fix the flow velocity to be $v_x=v_m\sin(2\pi x)$ with the amplitude 
$v_m=300$, which corresponds to $8.93\sqrt{k_BT_0/m_e}$. 
As discussed by \cite{KaufmanBlaes2016}, if the photon mean free path is much longer or smaller than the length scale of velocity gradient, which is the box size in this test, pure convergent/divergent flows cannot change the photon spectrum. The parameters in this test are chosen so that photon diffusion speed is comparable to the advection speed so that the spectrum can be significantly modified by bulk Compton. At the beginning when velocity is added, the isotropic lab frame specific intensities will no longer be isotropic in the co-moving frame. Scattering will tend to isotropize the intensities in the co-moving frame while thermal Compton as well as velocity gradient shift the photons in the frequency space. Spatial radiation pressure gradient also develops because of the non-uniform velocity gradient. We calculate spatial variation of the radiation energy density for each frequency group as
\begin{eqnarray}
	\delta E_r(x,\tnu)\equiv \frac{E_r(x,\tnu)-<E_r(x,\tnu)>_x}{<E_r(x,\tnu)>_x},
	\label{eq:delta_Er_x_nu}
\end{eqnarray}
where $<E_r(x,\tnu)>_x$ is the spatially averaged monochromatic radiation energy density at frequency $\tnu$. Spatial and frequency distribution of $\delta E_r(x,\tnu)$ at time $t=2$, which corresponds to $y=[c\rho\kappa_{\rm es}kT_0/(m_ec^2)]t=2.76$, is shown in Figure \ref{fig:bulk_compton_spatial}. Since bulk motions change photon energies at the rate proportional to the photon frequency \citep{KaufmanBlaes2016},  at low frequencies ($\tnu \lessapprox 1$), photon diffusion dominates and $E_{r,\tnu}$ is very uniform across the whole simulation box with $|\delta E_r(x,\tnu)|$ 
smaller than $10^{-3}$.  The small enhancement of $E_r(x,\tnu)$ correlates with the velocity magnitude, which represents the difference between lab frame and co-moving frame radiation energy densities since $(v_x/c)^2<1.3\times 10^{-3}$. At high frequencies ($\tnu > 1$), $\delta E_r(x,\tnu)$ then correlates with $dv_x/dx$ very well. It peaks at the center where velocity converges and the minimum is at the edge where velocity diverges. 

Even though the integrated $dv_x/dx$ over the whole box is 0, total radiation energy density in the simulation domain still increases because radiation pressure at $x=0.5$ is slightly larger than the radiation pressure at $x=0$.  Therefore the net radiation energy density increasing rate due to the radiation work term is also proportional to $v^2$ in this test. The thermal Compton term then transfers energy from photons to gas as we fix the gas temperature to $T=1$, which becomes smaller than the effective Compton temperature of photons.  At time $t=2$, total radiation energy density $E_r$ integrated over the whole simulation domain and all the frequency groups reaches $15 a_rT_0^4$ and spectrum of the box integrated $E_r(\tnu)$ is shown as the red line in Figure \ref{fig:bulk_compton_converge}. The peak has moved to higher frequencies compared with the initial blackbody spectrum. Since we only have scattering in this test and the simulation domain is periodic, we can compare the spectrum to the Bose-Einstein distribution $n(\tnu)=1/[\lambda \exp(\tnu/T_{\text{eff}})-1]$, where the effective temperature can be estimated as $T_{\text{eff}}=T+m_e v_m^2/(3k_B)=27.57T$ \citep{Psaltisetal1997,KaufmanBlaes2016} and $\lambda$ is related to the chemical potential of the distribution. We then fix the parameter $\lambda$ by requiring that the total radiation energy density from this distribution is the same as $E_r$. The spectrum is shown as the dashed green line in Figure \ref{fig:bulk_compton_converge}, which matches the peak of the numerical solution very well. But spectrum of the numerical solution is slightly flatter in both the low and high frequency parts, which is likely because this is a mixture of photons from different locations of the  box with different local spectra as shown in Figure \ref{fig:bulk_compton_spatial}.       

\begin{figure}[htp]
	\centering
	\includegraphics[width=1.0\hsize]{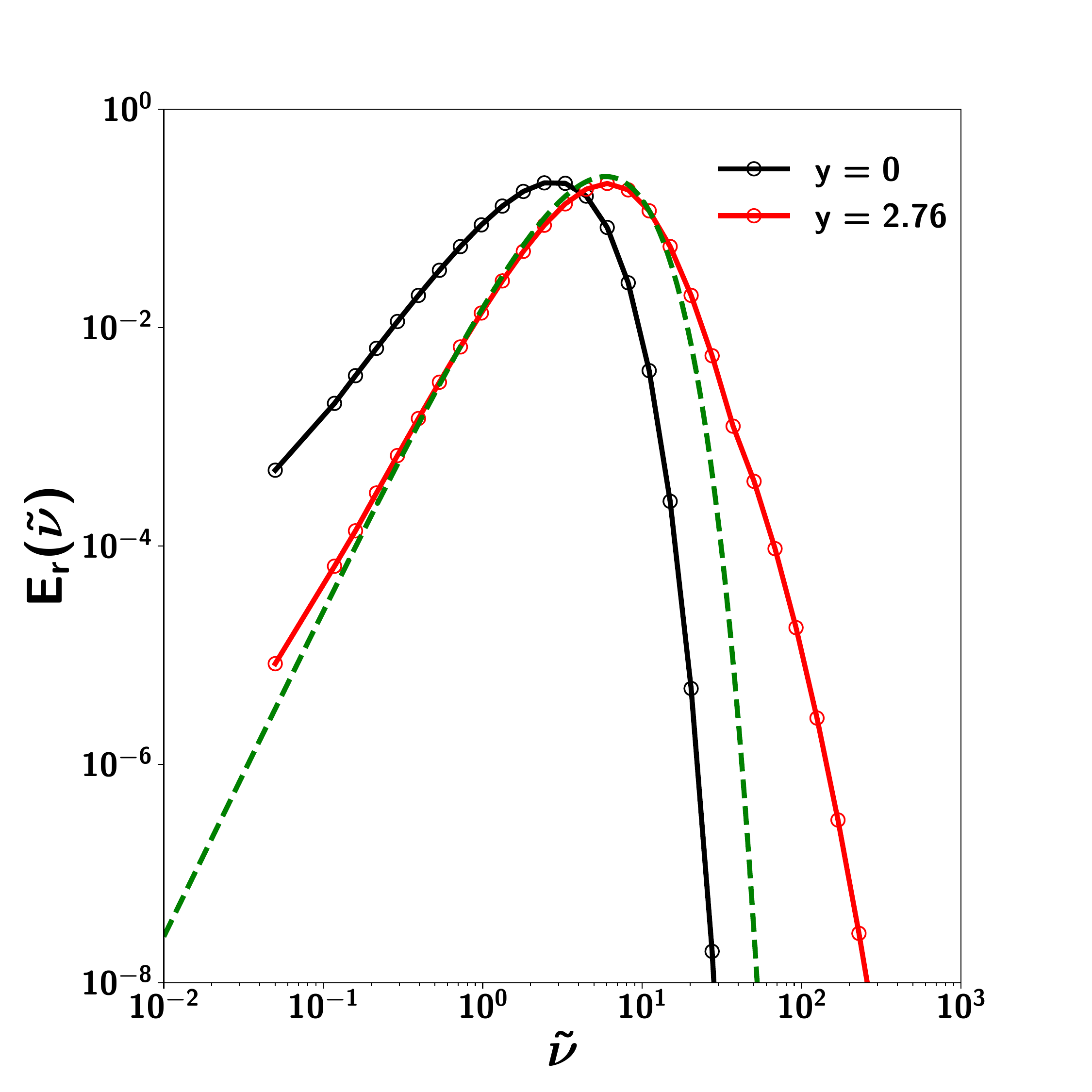}
	\caption{Spectrum of the spatially averaged monochromatic radiation energy density $E_r(\tnu)$ resulting from the bulk Compton of a shear flow as described in section \ref{sec:bulk_compton}. The black and red lines are for the initial blackbody spectrum and spectrum at time $t=2$ corresponding to the Compton $y$ parameter $2.76$. The dashed green line is a spectrum with Bose-Einstein distribution at the effective temperature $1.99T$ based on \cite{KaufmanBlaes2016}, where the gas temperature $T=1$ in our unit system. We fix the chemical potential in the distribution based on the total radiation energy density of the numerical solution at this time.  }
	\label{fig:bulk_compton_shear}
\end{figure}

We then consider a shear flow in the 2D domain $(x,y)\in[0,1]\times [0,1]$ as $v_y=v_m\sin(2\pi x)$ with $v_m=300$ and the horizontal velocity is 0.  We use $32\times 32$ grid points for the spatial resolution and 40 angles for the angular resolution. All the other quantities are the same as in the last test. Radiation energy density is increased everywhere initially via the radiation viscosity term. Then the thermal  Compton term transfers energy from photons to the gas. The energy increasing rate in this case is typically small because the off-diagonal components of radiation pressure is small compared with the diagonal components and only the velocity shear rate across the photon mean free path can contribute. 
Spectrum of spatially averaged radiation energy density at time $t=2$ is shown in Figure \ref{fig:bulk_compton_shear}. Total radiation energy density $E_r$ is only increased to $2.15a_rT_0^4$ at this time, which is $14\%$ of $E_r$ in the last case at the corresponding time. We estimate the effective temperature of photons resulting from this shear flow using equation 39 of \cite{KaufmanBlaes2016} as $T_{\text{eff}}=T+f(\tau_k)m_e \langle v^2\rangle /(3k_B)=1.99T$, where  the mean velocity square is $\langle v^2\rangle=v_m^2/2$ and the  optical depth reduction factor is $f_{\tau_k}=0.074$ for $\tau_k=10/(2\pi)$. A spectrum of Bose-Einstein distribution at the effective temperature $T_{\text{eff}}$ is shown as the dashed green line in Figure \ref{fig:bulk_compton_shear}. The Bose-Einstein distribution is also normalized to match $E_r$ in the numerical solution. It can explain the peak and low frequency end of the calculated spectrum very well. However, the numerical solution shows  a much more extended high frequency tail, which cannot be captured by this Bose-Einstein distribution.

\subsubsection{Compton Scattering in a Stratified Atmosphere}
\label{sec:compton_disk}

\begin{figure}[htp]
	\centering
	\includegraphics[width=1.0\hsize]{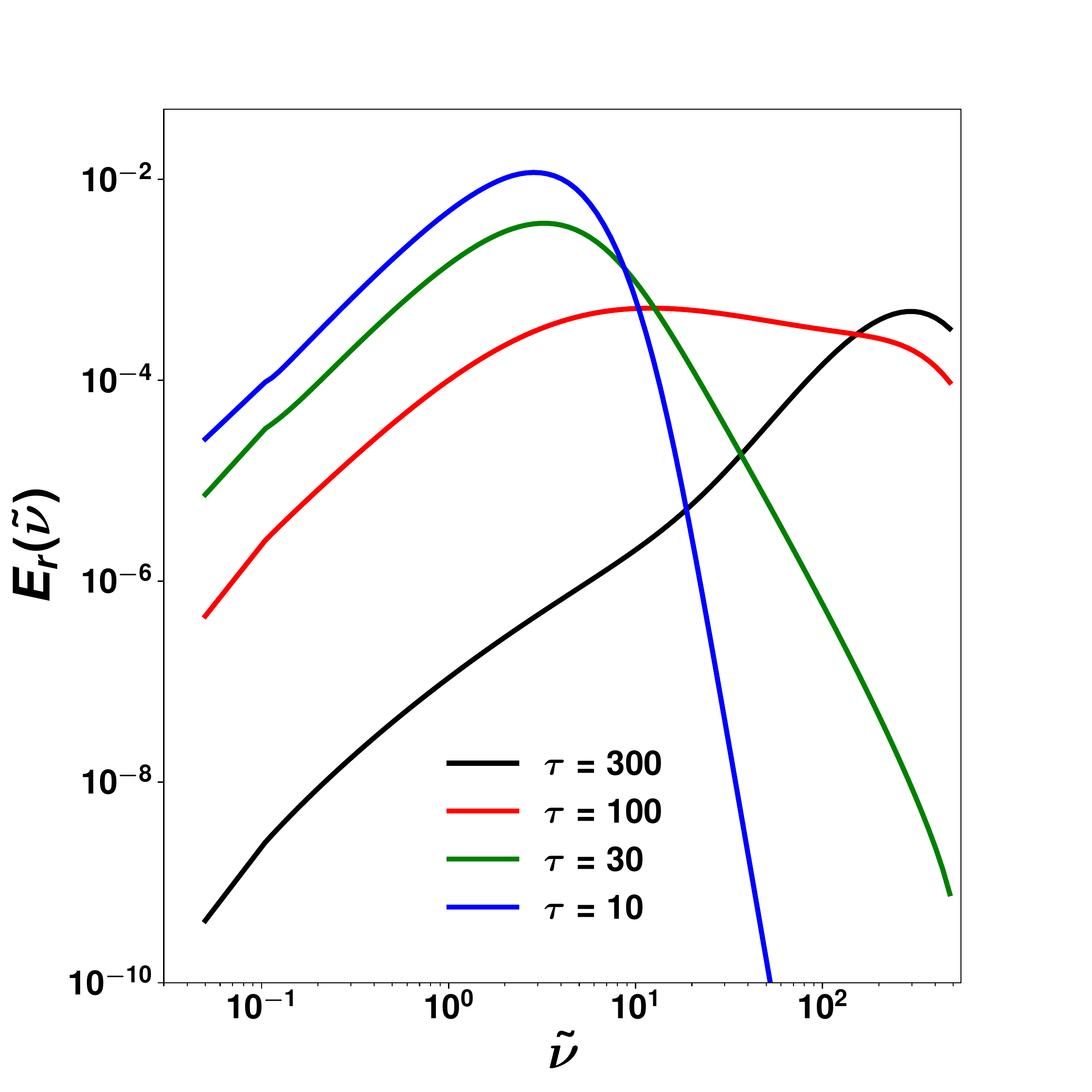}
	\caption{Spectra of the monochromatic radiation energy density from the surface of a column when total scattering optical depth varies from $300$ to $10$ as described in section \ref{sec:compton_disk}. The Compton 
	$y$ parameter varies from $15$ to $0.05$ correspondingly. }
	\label{fig:compton_disk}
\end{figure}

To check how the Kompaneets solver is coupled to other terms in the RT equation, we set up a 1D domain covering $x\in[0,3L_0]$  using 128 grid points with the density profile $\rho(x)=\rho_0 \exp(-x^2)$. Here the length unit is $L_0=3\times 10^{14}$ cm and $\rho_0=10^{-10}$ g/cm$^3$.  We vary the scattering opacity $\kappa_{\rm es}$ so that total optical depth $\tau=\int_0^{3L_0} \kappa_{\rm es}\rho dx$ reaches the desired value. Absorption opacity is 0 in this test. Gas temperature is set to a constant value $T=10^6$ K. We inject photons from the bottom at $x=0$ with a $10^4$ K black body spectrum isotropically. At the top boundary $x=3L_0$, we copy the outgoing specific intensities from the last active zone to the ghost zones and set the incoming specific intensities to be 0. We only use two angles per cell (the two-stream approximation) so that we can check the numerical solution easily. All the hydro variables are held fixed in this test. 
We use $100$ frequency bins and logarithmic frequency grid to cover $\tnu \in [0.1,500]$. We collect photons at the surface $x=3L_0$ to determine its spectra when we vary the total optical depth $\tau$ from 300 to 10,  which are shown in Figure \ref{fig:compton_disk}. All terms in the Kompaneets equation are included in this test. 
The typical diffusion time from the bottom to top is $t=3L_0\tau/c$. Then the Compton $y$ parameter is given by $y=\left[c\rho\kappa_{\rm es}k_BT/(m_ec^2)\right]3L_0\tau/c\approx 5\times 10^{-4}\tau^2$. When $\tau=30$ and $10$, $y\ll 1$ and the change of radiation spectra due to Compton scattering is minimal. Spectra of the emitting photons are close to the black body spectrum we injected from the base. For $\tau=100$, the spectrum is significantly modified by Compton scattering. While the whole domain has $\tau=300$ and $y\gg 1$, the spectrum is saturated to the gas temperature $T=10^6$ K. To check the numerical solution, we discretize the steady state transport equation \ref{eq:rad_mhd} and the Kompaneets equation in both  frequency and spatial grids. We confirm that they are indeed satisfied by the numerical solution. 

\begin{figure}[htp]
	\centering
	\includegraphics[width=1.0\hsize]{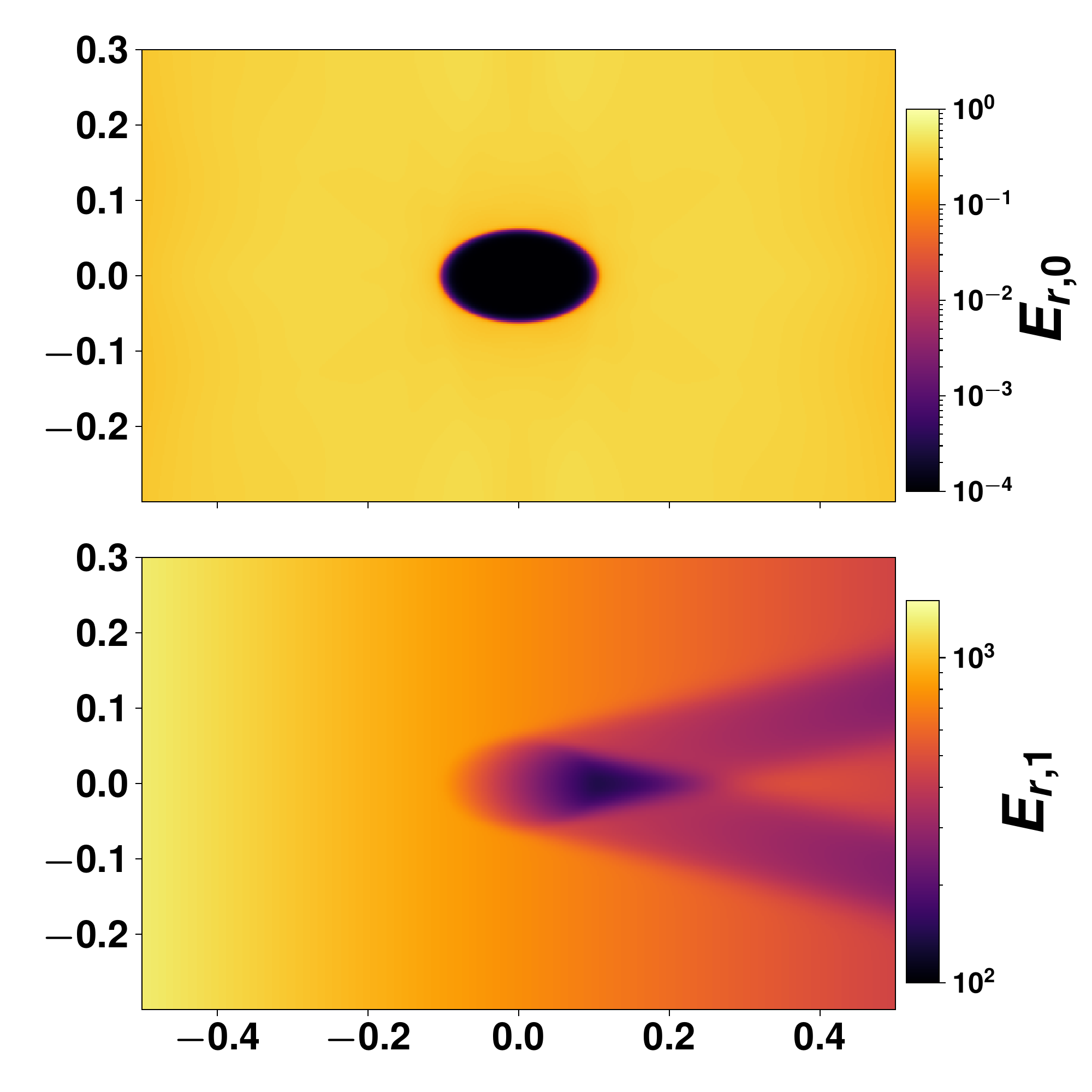}
	\caption{Radiation energy densities of the two frequency groups used in the shadow test described in section \ref{sec:shadow}. The top panel shows spatial distribution of radiation energy density $E_{r,0}$ in the first frequency group, which comes from thermal emission of the gas. The bottom panel shows the radiation energy density in the second frequency group  $E_{r,1}$, which is dominated by radiation coming from the left boundary and cast shadows behind the cloud.  }
	\label{fig:shadow}
\end{figure}

\subsection{Shadow Test with Two Frequency Groups}
\label{sec:shadow}
One application of the multi-group RT is that it can handle radiation coming from different sources, which may have quite different characteristic frequencies and associated opacities. As a prove of principle, we extend the classical shadow test  to two frequency groups, which cover the frequency range $\tnu\in [0,5)$ and $\tnu\in[5,\infty)$ respectively. We use a 2D simulation domain $(x,y)\in [-0.5,0.5]\times[-0.3,0.3]$ with $512\times 256$ grid points.  Each cell has 40 angles for specific intensities in this test. Temperature unit $T_0$ is chosen so that the dimensionless speed of light is $\Crat=1.9\times 10^5$. Density has the shape $\rho(x,y)=1+9/\left\{1+\exp[10\left[(x/0.1)^2+(y/0.06)^2-1\right]]\right\}$ and temperature is set to be $T(x,y)=1/\rho(x,y)$, which varies from $0.1$ inside the cloud to $1$ outside the cloud. Gas velocity is 0 and all gas quantities are fixed in this test. The first frequency group represents low energy photons emitted by the gas and it has absorption opacity $\kappa_{a,0}=\kappa_{p,0}=\rho T^{-3.5}$. Scattering opacity is 0 for this group. At the left and right boundaries, incoming specific intensities are set to be 0 while outgoing ones are copied from the last active zone to the ghost zones. The second frequency group represents high energy photons emitted by external sources. At the left boundary, incoming intensities with angle $18^{\circ}$ relative to the $x$ axis are set to $a_rT_i^4/(8\pi w_n)$ while all the other rays are set to be 0, where $T_i=6$ and $w_n$ is the angular quadrature weight for each ray. This is designed so that the mean radiation energy density for this group injected from the left boundary is $E_{r,1}=4\pi\sum_n I_f(n) w_n=a_rT_i^4$. At the right boundary, the same boundary condition as used for the first group is  adopted.  The top and bottom boundaries are periodic for both frequency groups.  The second group is set to have a constant absorption opacity $\kappa_{a,1}=\kappa_{p,1}=1$ and scattering opacity is also 0. Radiation energy densities in the two frequency groups after the system reaches steady state are shown in Figure \ref{fig:shadow}. For the first group, $E_{r,0}/a_rT_0^4$ is very close to $(T/T_0)^4$, which varies from $10^{-4}$  to $1$. Total optical depth across the cloud in the second frequency group is only $2.8$. Therefore, photons in this group penetrate the cloud while get attenuated. Umbra and penumbra are clearly captured in this frequency group, although size of the umbra is reduced due to lower optical depth compared with similar test done for the grey case \citep{Jiang2021}. This will also result in different dynamics of the cloud if we let gas evolve with radiation field together \citep{Progaetal2014}, which will be studied in the future with more realistic conditions.

\subsection{Bondi Accretion with Compton Scattering}
\label{sec:bondi}

To demonstrate that the algorithm described here is able to handle the full radiation hydrodynamics with Compton scattering together, we set up a spherically symmetric one dimensional (1D) Bondi accretion flow and see 
how the solution will evolve with the full radiation hydrodynamics turned on.  We do not include magnetic field here as it does not make any difference in 1D. There is no analytical solution to compare with in this case. Instead, this is commonly used as a proof of principle calculation for radiation MHD codes \citep{Fragileetal2012} as well as tests for Compton scattering \citep{Titarchuketal1997,Psaltisetal1997,Psaltis2001,Turollaetal2002}. 

\begin{figure}[htp]
	\centering
	\includegraphics[width=1.0\hsize]{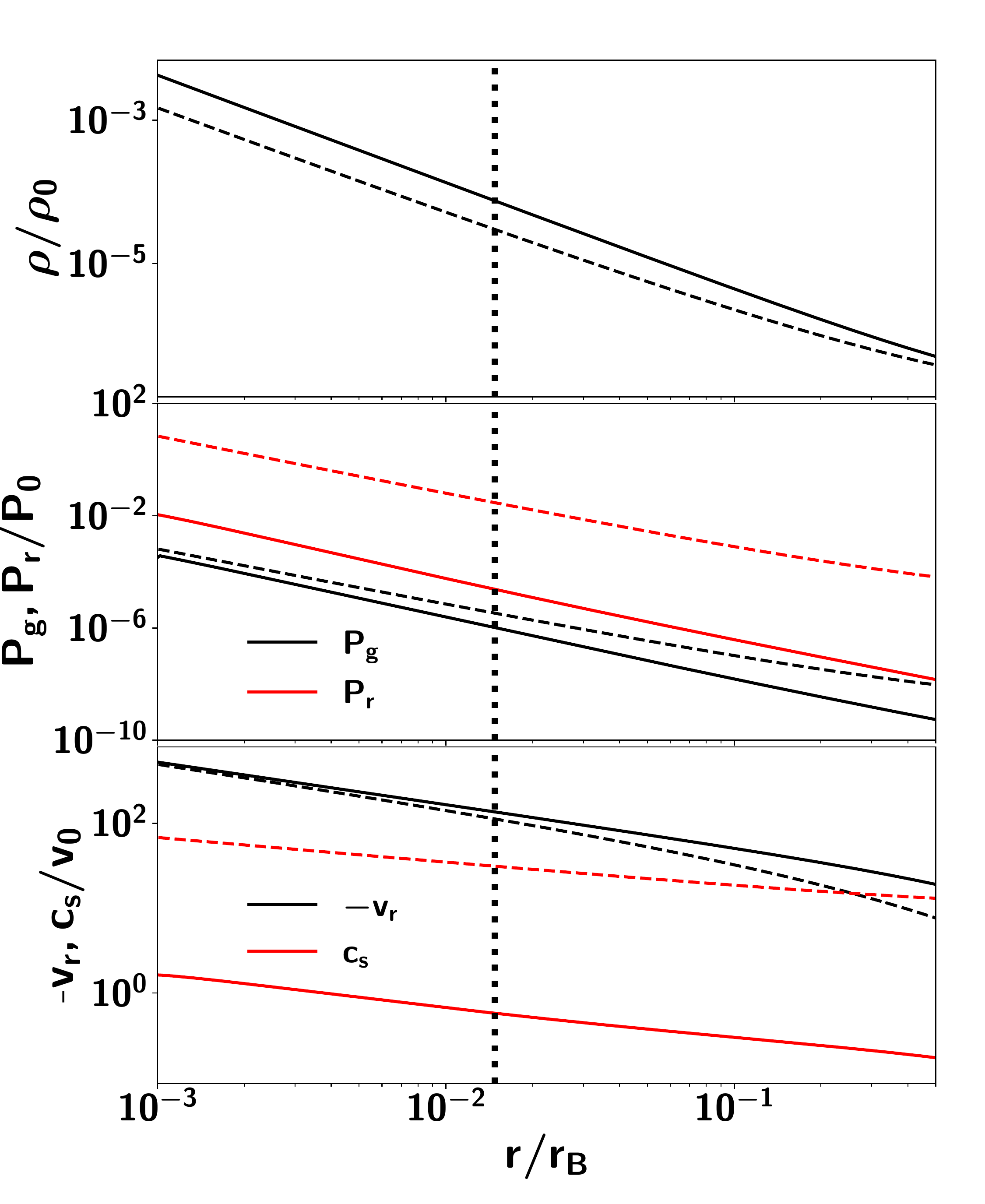}
	\caption{Radial profiles of density ($\rho$, top panel), gas ($P_g$) and radiation ($P_r$) pressure (middle panel), inflow velocity ($v_r$) and isothermal sound speed ($c_s$) as defined by the total pressure (bottom panel) for the Bondi type solution as described in Section \ref{sec:bondi}. The solid lines are for the steady state solution while the dashed lines are from the initial condition, which is calculated based on the classical Bondi solution. The dotted vertical line indicates the location of effective absorption photosphere for the frequency group centered at $\tnu=0.02$. }
	\label{fig:bondi1d}
\end{figure}

\begin{figure}[htp]
	\centering
	\includegraphics[width=1.0\hsize]{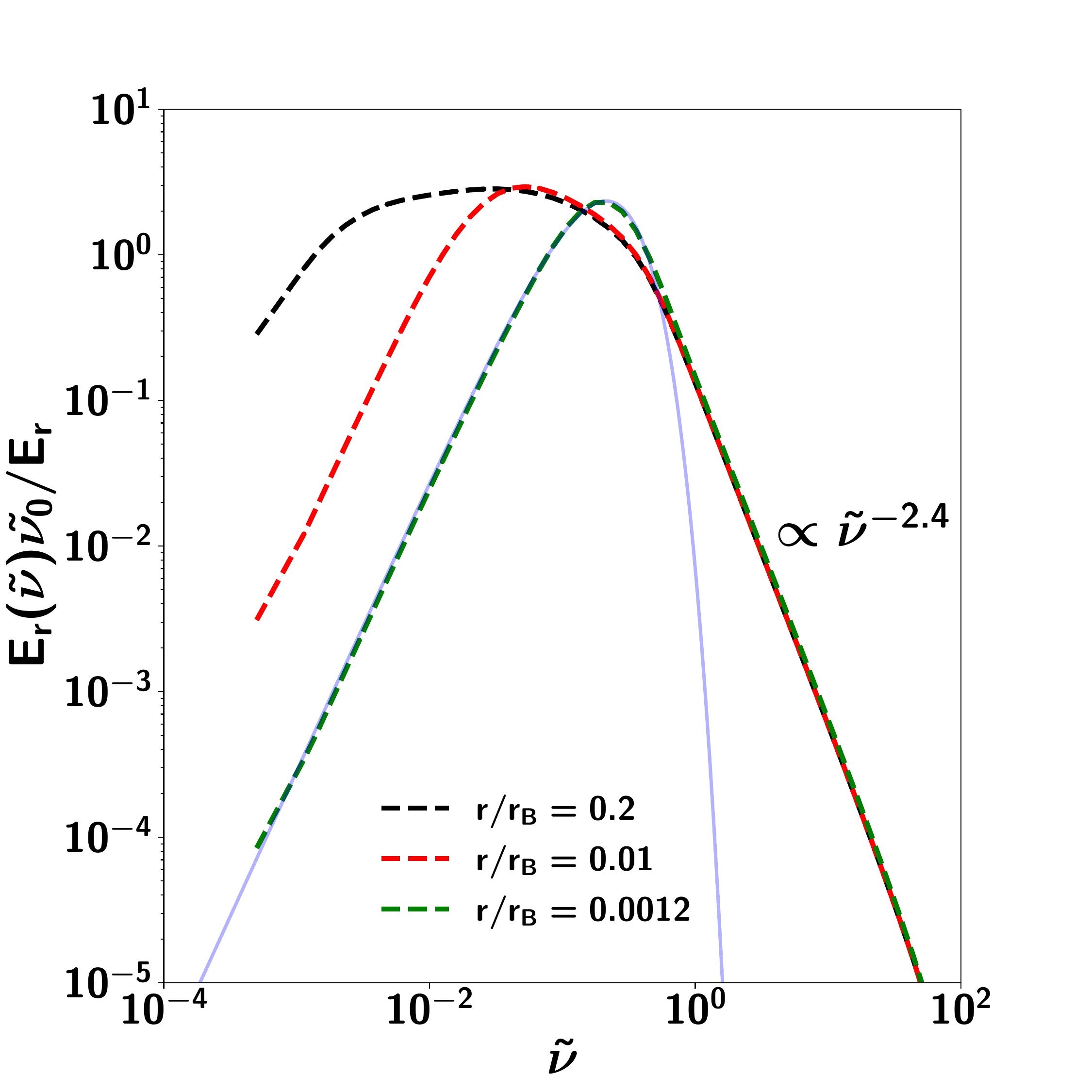}
	\caption{Spectrum of steady state Bondi type solution as described in Section \ref{sec:bondi} at radii $r/r_B=0.2$ (dashed black line), $r/r_B=0.01$ (dashed red line) and $r/r_B=0.0012$ (dashed green line). The monochromatic 
		radiation energy density $E_r(\tnu)$ is scaled with $E_r/\tnu_0$ at each radius. The solid blue line is a blackbody spectrum fitted to $E_r(\tnu)$ at $r/r_B=0.0012$. The spectra at frequencies $\tnu > 1$ can be fitted with a power law $\tnu^{-2.4}$.}
	\label{fig:bondi_spectrum}
\end{figure}

We initialize the simulation with the classical Bondi solution \citep{Bondi1952,Franketal2002} with an effective adiabatic index $\gamma_B=1.4$. We consider accretion onto a black hole with mass $\mbh=10^6M_{\odot}$, where $M_{\odot}$ is the solar mass.  Adiabatic sound speed and density at infinity are chosen to be $c_{\infty}=4.33\times 10^7\ \text{cm/s}$ and $\rho_{\infty}=10^{-17}\ \text{g/cm}^3$. Mass accretion rate of the Bondi solution is then $\dot{M}_B=\pi r_B^2\rho_{\infty}c_{\infty}\left(2/(5-3\gamma_B)\right)^{\left(5-3\gamma_B\right)/\left(2\left(\gamma_B-1\right)\right)}$, where Bondi radius $r_B$ is defined as $r_B=G\mbh/c_{\infty}^2$ and $G$ is the gravitational constant. This is about $12$ times the Eddington accretion rate for the parameters we choose. Since $r_B$ is $4.8\times 10^5$ times the gravitational radius, we cannot cover the whole radial range from the gravitational radius to $r_B$. Instead we focus on the outer region near $r_B$ to see how the solution will be modified with RT turned on. This will also cover the photosphere, which is not the case if only the free fall part in the Bondi solution is considered \citep{Titarchuketal1997,Fragileetal2012}.
We use 1D spherical polar coordinate covering radial range $r/r_B\in[10^{-3},1]$ with 512 grid points logarithmically. In the frequency space, we use logarithmic grid to cover $\tnu\in[10^{-3},10^2]$ with $N_f=50$.
We set density and pressure in the inner ghost zones so that their gradients are continuous across the inner boundary. Radial velocity in the inner ghost zones is set by requiring the mass flux to be continuous after density is determined in the boundary. The same boundary condition is also applied to the fluid variables at the outer boundary.  For specific intensities at both inner and outer boundaries, we require $I r^2$ to be continuous across the boundary for all angles and frequency groups. To initialize the simulation, we first calculate the radial profiles of the classical Bondi solution based on the fact that mass accretion rate and the Bernoulli parameter $v_r^2/2+\gamma_B c_s^2/\left(\gamma_B-1\right)-G\mbh/r$ are constant \citep{Franketal2002}.  Here $c_s^2$ is the ratio between total pressure and density. We then require that radiation and gas are in thermal equilibrium with the total pressure unchanged so that $P_g+P_r=\rho c_s^2$. Radiation pressure is larger than  gas pressure by a factor of $6.8\times 10^3-10^4$ but gas temperature only varies from $0.02T_0$ to $0.44T_0$, where $T_0=10^5$ K is the fiducial unit for temperature. The fiducial density and velocity units are chosen to be $\rho_0=10^{-10}\ \text{g/cm}^3,v_0=3.66\times 10^6 \text{cm/s}$ so that the dimensionless parameters $\Prat=565.2$ and $\Crat=8193.1$ for this test problem. 
We take the electron scattering opacity to be $0.34\ \text{cm}^2/\text{g}$ and frequency dependent absorption opacity as $3.68\times 10^{56} T_a^{-1/2}\rho\nu^{-3}  \text{cm}^2/\text{g}$, where the temperature parameter is $T_a\equiv \text{max}(T,0.1T_0)$. For each frequency group, we simply take the frequency value at the center of each bin to calculate the absorption opacity. 

Radial profiles of various quantities for the initial condition as well as steady state solution are shown in Figure \ref{fig:bondi1d}. Density and radial velocity increase slightly to make the accretion rate increased by a factor of 3. However, gas temperature has dropped by a factor of $5-50$ from the inner to outer regions, which makes the whole simulation domain become supersonic.  Radiation pressure is only $30$ times the gas pressure. This implies that the sonic point will be located at a much larger radius than what the original Bondi solution suggests. 
For the 1D accretion flow, the only dissipation mechanism is due to compression, which is very inefficient. For the radial range we have covered here, radiative luminosity in steady state is only $6.1\times 10^{-5}$ of the Eddington luminosity. 

The calculation also provides spectra of the radiation field directly, which are shown in Figure \ref{fig:bondi_spectrum} for three representative locations. We have normalized the radiation energy density in each frequency group by the frequency integrated radiation energy density at each radius so that we can compare their spectrum shape in the same plot. Total optical depth for scattering opacity across the simulation domain is $20$ for all frequency groups, while integrated optical depth for absorption varies from $10^5$ at the low frequency end to $10^{-9}$ at the high frequency group. Optical depth for the effective absorption, which is defined as the geometric mean of absorption and total opacity, is smaller than 1 for frequency groups with $\tnu > 0.28$. Near the inner region at $0.0012r_B$, spectrum in the low frequency range ($\tnu \lesssim 0.5$) follows the blackbody shape very well. For high frequencies ($\tnu \gtrsim 0.5$) where effective absorption optical depth is smaller than 1, the spectrum follows a power law shape $\tnu^{-2.4}$ very well. In fact, the high frequency tails of spectra at all three radii follow the same power law shape, which is likely produced by bulk Comptonization due to the convergent flow. The inflow velocity is larger than $\sqrt{3k_BT/m_e}$ by a factor of $10$ to $30$ in the whole simulation domain and gas temperature is too small to produce any thermal Compton effect. Since radial velocity ends up reaching the free fall limit in the whole simulation domain, if only first order $v/c$ effect is considered, the high frequency spectrum will have a universal power law shape $\tnu^{-2}$ \citep{PayneBlandford1981}. When all the terms are considered properly, the power law slope will depend on detailed properties of the flow, which can vary between $-2$ and $-3$ for the accretion rate we are considering here according to the model studied by \cite{Turollaetal2002}. This is very consistent with the power law slope we get. At larger radii, more frequency groups become optically thin for effective absorption and scattering optical depth is also reduced, while parts of the spectra at the lowest frequency end still follow the blackbody shape. Spectrum peaks move to lower frequencies due to smaller gas temperature. 
If we treat bulk Compton scattering as thermal Compton with effective temperature $T_{\text{eff}}=m_ev_r^2/(3k_B)$ \citep{KaufmanBlaes2016}, we can estimate the Compton $y$ parameter as $y=\tau^2\left(v_r/c\right)^2$, which is 1 at $r/r_B=0.0013$. This explains why the peak regions of the spectra become flatter at $r/r_B=0.01$ and $r/r_B=0.2$ similar to the spectra shown in Figure \ref{fig:compton_disk} (the red line).

\section{Discussion}
\label{sec:discuss}

The multi-group RT algorithm we have developed here is fully implicit for both the spatial transport term and the source terms. This is designed to work efficiently for a wide range of problems, particularly when the typical flow speed, sound speed 
or Alfv\'en velocity is much smaller than the speed of light. Therefore, time step of the whole radiation MHD scheme is not limited by the speed of light. The number of iterations we need for convergence in the implicit scheme is typically much smaller than the ratio between speed of light and the largest signal speed given by the MHD equations for all the test problems we have done. However, for systems with a flow speed comparable to the speed of light, it will be more efficient to solve the spatial transport term explicitly while still solve the source terms implicitly \citep{Jiangetal2014}. This can happen for either relativistic flows around compact objects, or non-relativistic systems with signal speed larger than $\gtrsim 0.1c$. 
In this way, time step will be limited by the CFL condition as defined by the speed of light but no iteration over the whole simulation domain is needed, as the implicit update for the source terms is completely local for each cell. This will also improve the parallel efficiency of the whole algorithm. It is very straightforward to convert the algorithm developed here to that case as we just need to change the term $\Delta t c\bn\cdot\bfnabla I_f^{m+1}$ in equation \ref{eq:discretized_equation} to $\Delta t c\bn\cdot\bfnabla I_f^{m}$. The way we solve all the other terms are unchanged.  

Performance of the overall multi-group radiation MHD algorithm will strongly depend on the number of iterations it takes to converge, which can vary significantly for different applications. The cost for each iteration is linearly proportional to $N\times N_f$. Coupling between different frequency groups in the source terms causes a few extra fourth order polynomial solvers (see section \ref{sec:solve_source}) compared with the grey case. However, this additional cost is a negligible fraction of the overall cost. As a reference of performance for real  applications that have used the code developed by \cite{Jiang2021}, the simulations described in \cite{Goldbergetal2022} adopt 120 angles per cell with $128^3$ spatial resolution and it takes about $10$ iterations to achieve a relative accuracy of $10^{-6}$ per time step. The code can update $10^4$ cells per second per core using more than $45$ skylake nodes with $40$ cores per node. To update $10^7$ cycles, which are typically needed to cover a few thermal time scales, it will cost about $10^6$ core hours. Adding $\approx 10$ frequency groups will likely increase the cost by an order of magnitude unless a fewer number of angles can be used. 

The scheme can also be used to calculate broad band spectrum when gas properties are given. For example, we can post process the simulation data generated by MHD simulations. For this purpose, we simply do not add radiation energy and momentum source terms back to the MHD equations. Even though the algorithm solves the full time \emph{dependent} RT equation, we can evolve the system to reach steady state to get the broadband spectrum of the flow. This behaves like a relaxation method for time \emph{independent} RT equation as normally assumed for spectrum calculations. However, future development is needed to speed up the convergence and the rate to reach steady state for this purpose. 
 
The current algorithm will need a lot of frequency groups to follow the Doppler shift of each individual line in order to model line transport in principle, which is not affordable in real applications. We also assume the frequency grid is the same at different spatial locations for simplicity. This is not optimal when features in the frequency space shift with space. This can happen when temperature varies significantly with space, or Doppler effect causes systematic shift of frequencies. Adaptive frequency grid can be used in this case, which means we can use different frequency grids at different spatial locations. This is also necessary to model the line transport efficiently. It will require mapping specific intensities in different frequency grids at the boundary of each cell, which will be developed in future work. 

Solvers for frequency and angular resolved RT calculations coupled with MHD simulations have been implemented in various codes to model stellar atmosphere and convection. The STAGGER code \citep{NordlundGalsgaard1995,Sulisetal2020} uses the Feautrier's scheme to solve the RT equation along several inclined rays (one vertical, eight inclined) while  MURaM \citep{Vogleretal2005} and CO5BOLD \citep{Freytagetal2012} adopt the short characteristic approach. All these codes bin the monochromatic opacities into several groups according to the height at which they mainly contribute to the radiative heating rate\citep{Nordlund1982,Vogler2004}. The opacity groups may not be continuous in the frequency space. Our algorithm differs from these codes in many ways. They all neglect the time dependence as well as velocity dependence of specific intensities, which will significantly complicate any kind of opacity binning scheme. Radiation field is used to determine the heating and cooling rate of gas while momentum exchange between photons and gas is not considered in these codes. Simulations using these codes are typically done in Cartesian coordinate as short characteristic approach is not easy to implement in general curvilinear coordinate systems. 

The algorithm developed here opens up new opportunities for a wide range of applications. Just to mention a few examples here, it can be used to improve radiation MHD simulations of black hole accretion disks \citep{Jiangetal2019} with a much better solver for Compton scattering in the corona region. It will also directly produce broad band spectra without the need of post processing. The scheme is also able to capture the collapse of hot accretion flows from optically thin to optically thick regimes due to Compton cooling much more efficiently compared with Monte Carlo method \citep{Dexteretal2021}. Interactions between supernova shocks and circumstellar material \citep{Margalitetal2022} can also be simulated accurately with this scheme. 
More importantly, the algorithm can be used to quantify whether grey approximation is appropriate or not, which is barely demonstrated before. When the dominant opacity depends on frequency, the algorithm can be used to test the accuracy of commonly adopted Planck mean and Rosseland mean opacities for thermal and momentum couplings between radiation and gas. A simple example is free-free opacity, which varies with frequency as $\nu^{-3}\left[1-\exp(-h\nu/k_BT)\right]$ for a given density and temperature and it is very relevant in many astrophysical systems \citep{RybickiLightman1986}. It can be easily shown that the Rosseland mean value is smaller than the Planck mean value by a factor of $\approx 30$. The actual opacity that determines the radiation force for a given luminosity will depend on the full radiation spectrum and it is likely between the Rosseland mean and Planck mean values. 
Rosseland mean opacity only applies when it is optically thick for the whole frequency range and radiation and gas are in thermal equilibrium. Similar conclusion has been found for 
Wolf-Rayet stars where the Rosseland mean opacity significantly underestimates the flux weighted value based on frequency dependent RT calculations \citep{Sanderetal2020}. 
We can resolve the frequency dependence of free-free opacity very well using $10$ to $20$ frequency groups that covers $\approx 0.1h\nu/k_BT$ to $\approx 10 h\nu/k_BT$ logarithmically, in which case the Rosseland and Planck mean values defined in each group will be very close to each other. The algorithm developed here will also be more accurate than many other multi-group approaches that adopt diffusion like approximations since we also resolve the full angular dependence of the radiation field. The region where grey approximation fails is also likely the place where diffusion approximation cannot apply.

\section*{Acknowledgements}
The author thanks Omer Blaes for valuable discussions on Compton scattering, 
as well as the anonymous referee for helpful comments that improved the paper. 
Part of the work is done when the author was attending the Binary22 program in KITP, 
which was supported in part by the National Science Foundation under Grant No. NSF PHY-1748958. 
This work is also part of the TCAN collaboration supported by the grant 80NSSC21K0496.
The Center for Computational Astrophysics at the Flatiron Institute 
is supported by the Simons Foundation.

The simulations use the public available code {\sf Athena++} \citep{Stoneetal2020}. The analysis made significant use of the following packages: NumPy \citep{Harrisetal2020}, SciPy \citep{Virtanenetal2020}, and matplotlib \citep{Hunter2007}.

\bibliographystyle{aasjournal}
\bibliography{citations}

\end{CJK*}

\end{document}